\documentclass[acmsmall,screen,nonacm]{acmart}
\AtBeginDocument{%
  }

\usepackage{subfiles}
\usepackage{xspace}
\usepackage{subcaption}
\usepackage{color, colortbl}
\usepackage{multirow}
\usepackage{hhline}
\usepackage{framed}
\usepackage{listings}
\usepackage{amsmath,amsfonts}
\usepackage{pifont}
\usepackage{array}
\usepackage{enumitem}
\usepackage{pgfplots}
\usepackage{pgfplotstable}
\usepackage{csvsimple}

\usepackage{tikz}
\usetikzlibrary{matrix}
\usetikzlibrary{positioning}
\usetikzlibrary{shapes}
\usetikzlibrary{calc}
\usetikzlibrary{fit}

\pgfplotsset{compat=1.18}

\lstdefinestyle{CodeStyle}{
    commentstyle=\color{black},
    keywordstyle=\color{black},
    numberstyle=\tiny,
    stringstyle=\color{black},
    basicstyle=\linespread{0.8}\ttfamily\footnotesize,
    breakatwhitespace=false,         
    breaklines=true,                 
    captionpos=b,                    
    keepspaces=true,                 
    numbers=left,
    xleftmargin=1em,
    numbersep=5pt,                  
    showspaces=false,                
    showstringspaces=false,
    showtabs=false,                  
    tabsize=2,
    language=C
}

\newcolumntype{C}[1]{>{\centering\let\newline\\\arraybackslash\hspace{0pt}}m{#1}}

\begin{document}

\title{Agentic Code Optimization via Compiler-LLM Cooperation}

% \begin{comment}
\author{Benjamin Mikek}
\affiliation{%
  \institution{AWS AI}
  \city{New York}
  \state{NY}
  \country{USA}}
\affiliation{%
  \institution{Georgia Institute of Technology}
  \city{Atlanta}
  \state{GA}
  \country{USA}}
\email{bmikek@gatech.edu}

\author{Danylo Vashchilenko}
\affiliation{%
  \institution{AWS AI}
  \city{New York}
  \state{NY}
  \country{USA}}
  \authornote{Corresponding author}
\email{vdanylo@amazon.com}

\author{Bryan Lu}
\affiliation{%
  \institution{AWS AI}
  \city{Santa Clara}
  \state{CA}
  \country{USA}}
\email{yuzhelu@amazon.com}

\author{Panpan Xu}
\affiliation{%
  \institution{AWS AI}
  \city{Santa Clara}
  \state{CA}
  \country{USA}}
\email{xupanpan@amazon.com}

% \end{comment}


%\newcommand{\note}[2]{}
\newcommand{\note}[2]{\textbf{[#1: #2]}}
\newcommand{\bm}[1]{{\color{orange} \note{BM}{#1}}}

\newcommand{\ie}{\textit{i.e.}\xspace}
\newcommand{\codenet}{CodeNet\xspace}
\newcommand{\sumpara}[1]{\vspace{0.1cm}\noindent\textbf{#1.}\hspace{1mm}}

\newcommand{\tool}{ACCLAIM\xspace}
\newcommand{\toolst}{ACCLAIM*}
\newcommand{\toollongname}{\textbf{A}gentic \textbf{C}ooperation between \textbf{C}ompiler and \textbf{L}LM for \textbf{A}utomated \textbf{IM}provement of programs\xspace} %Please add any other suggestions!

\newcommand{\codein}[1]{{\footnotesize\tt{#1}}}

\newcommand{\cmark}{\ding{51}}%
\newcommand{\xmark}{\ding{55}}%

\newcommand{\rows}{}
\newcommand{\row}[1]{%
    \ifdefempty{\rows}
        {\xappto\rows{#1}}
        {\xappto\rows{,#1}}
}


\addtolength{\textfloatsep}{-.2in}
\addtolength{\abovecaptionskip}{-0.03in}
\addtolength{\belowcaptionskip}{-0.01in}
\addtolength{\floatsep}{-0.03in}

\begin{abstract}
Generating performant executables from high level languages is critical to software performance across a wide range of domains. Modern compilers perform this task by passing code through a series of well-studied optimizations at progressively lower levels of abstraction, but may miss optimization opportunities that require high-level reasoning about a program's purpose. Recent work has proposed using LLMs to fill this gap. While LLMs can achieve large speedups on some programs, they frequently generate code that is incorrect. In this work, we propose a method to balance the correctness of conventional compiler optimizations with the ``creativity'' of LLM-based code generation: compiler-LLM cooperation. Our approach integrates existing compiler optimization passes with LLM-based code generation at multiple levels of abstraction, retaining the best features of both types of code optimization. We realize our approach with a multi-agent system that includes (1) LLM-based optimization agents for each level of abstraction, (2) individual compiler constituents as tools, (3) an LLM-based test generation agent that probes the correctness and performance of generated code, and (4) a guiding LLM that orchestrates the other components. The strategy enables LLM-based optimization of input programs at multiple levels of abstraction and introduces a method for distributing computational budget between levels. Our extensive evaluation shows that compiler-LLM cooperation outperforms both existing compiler optimizations and level-specific LLM-based baselines, producing speedups up to $1.25\times$.
\end{abstract}

%The challenge is that while LLMs’ less conservative approach can introduce novel optimizations, it struggles with the precision required in low-level languages, and often requires large sets of representative training data. 

\maketitle

\section{Introduction}
\label{sec:introduction}
%Add sentence for why outliers are still worthwhile

Given an algorithm in a high-level language like C or PyTorch, the role of a compiler is to generate low-level executable code which is correct and efficient. Modern compilers~\cite{llvm} achieve both of these goals by analyzing and applying optimizations at several \textit{levels of abstraction}; for example, by processing from source code, through an intermediate representation (IR), and finally an assembly language like x86. Some compiler components move code from one abstraction level to the next (\textit{lowering}), while optimization passes improve a program without changing its level (\textit{rewriting}). Compilers excel at performing optimizations that are structured, deterministic, and general, like peephole optimizations, dead code elimination, and vectorization. Decades of research on verification~\cite{compcert, alive2} and testing~\cite{emi} have made modern compilers robust and driven down rates of miscompilation to negligible levels for most applications. But since compiled programs may be run billions of times, even small improvements over existing compilers can have large payoffs.

\begin{figure}
    \centering

    \begin{subfigure}{0.45\textwidth}
    \begin{tikzpicture}[node distance=2cm]
    % Define nodes
    \node[rectangle, align=center, fill=gray!50] (source) at (0,3.4) {Input \\ Source};
    \node[draw, ellipse, align=center, font={\tiny}] (frontend) at (0,2.2) {Compiler \\ Frontend};
    
    %\node[draw, rectangle] (ir) at (0,0) {IR};
    \node[draw, ellipse, align=center, font={\tiny}] (middleend) at (0,1.1) {Compiler \\ Middle-end};

    \node[draw, ellipse, align=center, font={\tiny}] (backend) at (0,0) {Compiler \\ Backend};
    
    \node[draw, rectangle] (assembly) at (0,-1) {Assembly};

    \node[cloud, aspect=2, fill={gray!50}] (training) at (-2.7,1.4) {Training Data};

    \node[draw, ellipse, align=center] (levelAgent) at (-2.7,-1) {Optimization \\ Agent};

    %\node[draw, ellipse, align=center, font = {\tiny}] (result) at (-1.4,-2) { Execution \\  Result};

    \node[rectangle, align=center, fill=gray!50] (output) at (-2.7,-2.4) {Output \\ Assembly};
    
    % Draw arrows
    \draw[->] (source) -- (frontend);
    \draw[->] (frontend) -- (middleend);
    \draw[->] (middleend) -- (backend);
    \draw[->] (backend) -- (assembly);
    %\draw[->] (ir) -- (assembly);
    \draw[->] (levelAgent) to [bend left=20] (assembly);
    \draw[->] (assembly) to [bend left=20] (levelAgent);
    %\draw[->] (assembly) to [bend left=20] (result);
    %\draw[->] (result) to [bend left=20] (levelAgent);
    \draw[->] (levelAgent) -- (output);

    \draw[->, draw=gray!50, line width=1mm] (training) to (levelAgent);

    % Style nodes as ellipses
    %\tikzstyle{every node}=[ellipse, draw, minimum width=2cm, minimum height=1cm]
\end{tikzpicture}
    \caption{Existing approaches to LLM-based code optimization, focusing on one level of abstraction.}
    \label{fig:intro-existing}
\end{subfigure}
\hfill
\begin{subfigure}{0.45\textwidth}
    \begin{tikzpicture}[node distance=2cm]
    % Define nodes
    
    %\node[draw, ellipse] (assembly) at (0,-1) {Assembly};

    %\node[draw, ellipse, font = {\small}] (source) at (0.8,2) {Source};
    \node[draw, ellipse, align=center, font = {\small}] (sourceAgent) at (1.6,1.4) {Source \\ Agent};
    %\draw[->] (sourceAgent) to [bend left=20] (source);
    %\draw[->] (source) to [bend left=20] (sourceAgent);

    %\node[draw, ellipse, font = {\small}] (ir) at (1.8,0) {IR};
    \node[draw, ellipse, align=center, font = {\small}] (irAgent) at (2.4,0) {IR \\ Agent};
    %\draw[->] (irAgent) to [bend left=20] (ir);
    %\draw[->] (ir) to [bend left=20] (irAgent);

    %\node[draw, ellipse, font = {\small}] (assembly) at (1.2,-2) {Assembly};
    \node[draw, ellipse, align=center, font = {\small}] (assemblyAgent) at (1.6,-1.4) {Assembly \\ Agent};
    %\draw[->] (assemblyAgent) to [bend left=20] (assembly);
    %\draw[->] (assembly) to [bend left=20] (assemblyAgent);

    \node[rectangle, align=center, fill=gray!50] (source) at (0,2.4) {Input \\ Source};

    \node[rectangle, align=center, fill=gray!50] (assembly) at (0,-2.4) {Output \\ Assembly};

    \node[draw, ellipse, align=center, font = {\small}] (frontend) at (-1.6,1.4) {Compiler \\ Frontend};

    \node[draw, ellipse, align=center, font = {\small}] (middleend) at (-2.4,0) {Compiler \\ Middle-end};

    \node[draw, ellipse, align=center, font = {\small}] (backend) at (-1.6,-1.4) {Compiler \\ Backend};

    %\node[draw, rectangle, font = {\tiny}, align=center] (frontend) at (0,1) {Compiler \\ Frontend};
    %\node[draw, rectangle, font = {\tiny}, align=center] (backend) at (0,-1) {Compiler \\ Backend};

    \node[draw, ellipse, align=center] (planner) at (0,0) {Guiding \\ Agent};

    \draw[->] (planner) to [bend left=10] (sourceAgent);
    \draw[->] (sourceAgent) to [bend left=10] (planner);

    \draw[->] (planner) to [bend left=10] (irAgent);
    \draw[->] (irAgent) to [bend left=10] (planner);

    \draw[->] (planner) to [bend left=10] (assemblyAgent);
    \draw[->] (assemblyAgent) to [bend left=10] (planner);

    \draw[->] (planner) to [bend left=10] (frontend);
    \draw[->] (frontend) to [bend left=10] (planner);

    \draw[->] (planner) to [bend left=10] (middleend);
    \draw[->] (middleend) to [bend left=10] (planner);

    \draw[->] (planner) to [bend left=10] (backend);
    \draw[->] (backend) to [bend left=10] (planner);

    \draw[->] (source) -- (planner);
    \draw[->] (planner) -- (assembly);

    %\node[draw, ellipse, align=center, font = {\tiny}] (result) at (-1.4,-2) { Execution \\  Result};
    
    % Draw arrows
    %\draw[->] (source) -- (ir);
    %\draw[->] (ir) -- (assembly);
    %\draw[->] (levelAgent) to [bend left=20] (assembly);
    %\draw[->] (assembly) to [bend left=20] (result);
    %\draw[->] (result) to [bend left=20] (levelAgent);

    % Style nodes as ellipses
    \tikzstyle{every node}=[ellipse, draw, minimum width=2cm, minimum height=1cm]
\end{tikzpicture}
    \caption{Our approach, using a guiding agent, a testing agent, and level-specific optimization agents at each level of abstraction.}
    \label{fig:intro-contribution}
\end{subfigure}

\caption{Comparison between existing work (left) and our multi-level approach (right).}
\label{fig:intro}
\end{figure}
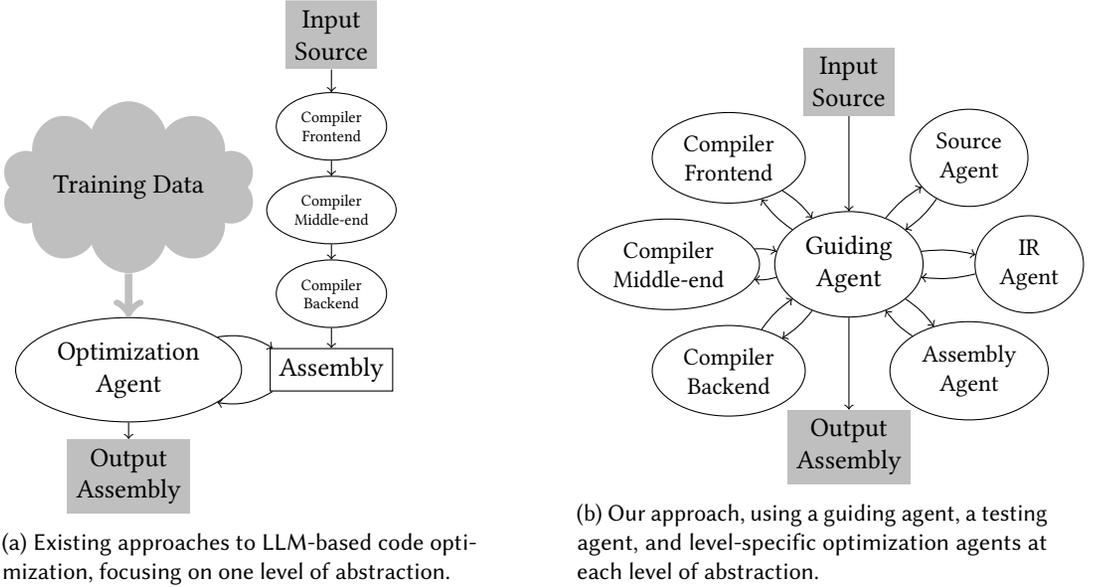

Recently, attention has turned to harnessing the power of large language models (LLMs) to augment this optimization process~\cite{searchBased, kernelbench, learningEdits}. A typical workflow for LLM-based optimization~\cite{assembly} is shown in Figure~\ref{fig:intro-existing}. An input program is first compiled to a particular level of abstraction, x86 assembly in this case. An LLM is then provided a context window containing the code and an optimization prompt; it produces optimized output code. LLM-generate code is often improved by iterative refinement, in which the LLM is provided feedback on the generated code and tries to improve on its own prior generation. The code generation model itself may be improved when more data is available via reinforcement learning, supervised fine-tuning, etc. LLM-based code optimization has shown promise for improving program performance at assembly level~\cite{assembly}, at source level~\cite{searchBased, learningEdits}, and in the GPU context~\cite{kernelbench, kevin, metrKernel}.

%In existing methodologies, the context window includes a program at a fixed level of abstraction and some optimization instructions. A language model is then tasked

%with rewriting at that level: producing a new program that has better performance than the original.  LLMs' context and reasoning abilities enable them to occasionally identify new types of optimizations that compilers miss---for instance, noticing that a loop can be replaced with a mathematical function. 

The downside of LLM-based code optimization is that models frequently generate incorrect code---rates of incorrect generation range from 10\% to as high as 90\% depending on model and prompting~\cite{learningEdits}. The problem is exacerbated at lower levels of abstraction, as LLMs struggle to comply with the precise syntax and correctness demands of low-level languages like assembly. \citet{assembly}, for example, report the best un-augmented model produces incorrect code 42\% of the time when rewriting, and almost always fails when generating assembly code from source. Worse performance at lower levels of abstraction occurs, in part, because model training datasets focus on high-level languages: assembly code makes up just 0.08\% of the popular code dataset The Stack, while C and C++ programs make up 13\%~\cite{the-stack}. Compiler IRs do not appear at all. Existing work combats low rates of correctness by utilizing in-context learning (ICL) with retrieved samples~\cite{kevin} and training methods like reinforcement learning (RL)~\cite{assembly, rlef} and supervised fine-tuning (SFT)~\cite{ecco, r1-interpreter}. These approaches, however, rely on the availability of curated prompts or training data---they trade better correctness for the availability of data. There is therefore a trade-off between LLMs and compilers. LLMs are more ``creative'' and can reason about more complex optimization opportunities than compilers in some instances, but often produce incorrect code---boosting correctness requires training or sampling data. Compilers on the other hand are more conservative, producing code that is correct every time but sometimes missing high-level optimization opportunities. 

In this work, we propose a method to obtain the best of both worlds: compiler-LLM cooperation. Our approach balances the optimization promise of LLMs with the conservative but correct approach of compilers by allowing the two to cooperate. The key enabling insight is that LLM-based optimization can be carried out not just at a single level of abstraction, but interleaved with calls to an existing compiler across all levels of abstraction. We then introduce a novel method of choosing how best to interleave LLM-based code generation with compiler components. Our approach, shown in Figure~\ref{fig:intro-contribution}, is to construct a multi-agent system that includes core LLM agents that optimize at each level of abstraction and a guiding agent that chooses when to use a compiler and when to use LLM-based optimization. Our realization of compiler-LLM cooperation does not rely on the availability of training data, but each level-specific agent is compatible with parallel sampling~\cite{sampling}, feedback loops~\cite{kevin}, and other training- and test-time methods of tuning.

Conceptually, our approach has three main advantages. First, giving a multi-agent system the freedom to choose how to interleave existing compiler calls and LLM-based optimizations balances LLMs' optimization strengths with the guaranteed correctness of compilers. Second, it enables LLM-based optimization at multiple levels of abstraction, allowing a single program to benefit from multiple classes of optimization and potentially unlocking synergies among optimizations at multiple levels of abstraction. Third, the compiler-LLM cooperation strategy is useful for cases in which large training sets are not available---while our approach is \textit{compatible} with strategies like SFT and RL, it can still produce correct code without access to data beyond the target program.

We realize compiler-LLM cooperation as \tool
(\toollongname), a multi-agent system that optimizes C programs targeting CPUs. Our architecture, shown in Figure~\ref{fig:intro}, consists of the following:
\begin{itemize}
    \item For each level of abstraction, a level-specific agent that performs LLM-based rewriting.
    \item A testing agent that generates test cases and provides correctness and performance feedback.
    \item A guiding agent with access to each level-specific agent, the testing agent, and the lowering and rewriting components of an existing compiler as tools.
\end{itemize}
Given an input program, the guiding agent is empowered to use reasoning tokens to determine what types of optimizations might be a good match. It can call the existing compiler components, level-specific agents, and the testing agent to optimize an input program, and use feedback to inform better optimization decisions on the fly. The output is a low-level program that, in the ideal case, performs better than an assembly program generated by a conventional compiler alone.

We evaluate \tool on a standard set of C programs and find that it achieves mean speedups of $1.25\times$ against \texttt{clang~-O3}, outperforming both level-specific and naive multi-level baselines with equal computation budgets. Our analysis of \tool's behavior shows that it makes use of all compiler components and core agents and frequently backtracks or re-calls agents when results are not satisfactory. In summary, this paper makes the following primary contributions:
\begin{itemize}
    \item We introduce a framework for LLM-based optimization across multiple levels of abstraction in a compiler pipeline.
    \item We develop an agentic workflow for dynamically choosing when and at which levels of abstraction to apply LLM-based optimization with an orchestration agent.
    \item We show that, combined, the multi-level approach and orchestration strategy outperform both single-level and naive multi-level baselines.
\end{itemize}

In the rest of this paper, Section~\ref{sec:motivating} gives an example to motivate multi-level LLM-based optimization. Section~\ref{sec:approach} describes our approach in detail, and Section~\ref{sec:evaluation} describes our realization of compiler-LLM cooperation and evaluates its effectiveness. Section~\ref{sec:discussion} discusses the limitations and implications of our results, while Section~\ref{sec:related} surveys related work and Section~\ref{sec:conclusion} concludes.

\section{Motivating Example}
\label{sec:motivating}
\begin{figure}
\begin{subfigure}{\textwidth}
\begin{lstlisting}[style=CodeStyle,escapechar=|]
int total = 0;
for (int i = 1; i <= k; i++) {
    int x = i;
    int res = 0;
    while (x > 0) {
        res += x & 1;
        x >>= 1;
    }
    total += res;
}
\end{lstlisting}
\caption{Original code on input $k$, containing two nested loops. The result is stored in \texttt{total}.}
\label{fig:mot-f}
\end{subfigure}
\begin{subfigure}{\textwidth}
\begin{lstlisting}[style=CodeStyle,escapechar=|]
while (remaining > 0) {
    total += (k / (bit_pos * 2)) * bit_pos;
    unsigned long remainder = k % (bit_pos * 2);
    if (remainder >= bit_pos) {
        total += remainder - bit_pos + 1;
    }
    bit_pos <<= 1;
    if (bit_pos > k) break;
    remaining >>= 1;
}

\end{lstlisting}
\caption{For loop after LLM-based optimization at source level.}
\label{fig:mot-llm}
\end{subfigure}

\begin{subfigure}{\textwidth}
\begin{lstlisting}[style=CodeStyle,escapechar=|]
%18 = load i32, ptr %5, align 4
%19 = and i32 %18, 1
%20 = load i32, ptr %6, align 4
%21 = add nsw i32 %20, %19
store i32 %21, ptr %6, align 4
%22 = load i32, ptr %5, align 4
%23 = ashr i32 %22, 1
store i32 %23, ptr %5, align 4
\end{lstlisting}
\caption{LLVM IR program implementing the inner loop from $f$.}
\label{fig:mot-ir}
\end{subfigure}

\begin{subfigure}{\textwidth}
\begin{lstlisting}[style=CodeStyle,escapechar=|]
%5 = phi i64 [ 1, %3 ], [ %11, %10 ]
%6 = phi i32 [ 0, %3 ], [ %9, %10 ]
%7 = trunc i64 %5 to i32
%8 = tail call i32 @llvm.ctpop.i32(i32 %7)
%9 = add i32 %6, %8
\end{lstlisting}
\caption{LLVM IR program after LLM-based optimization.}
\label{fig:mot-llmir}
\end{subfigure}

\begin{subfigure}{\textwidth}
\begin{lstlisting}[style=CodeStyle,escapechar=|]
call <4 x i32> @llvm.ctpop.v4i32(<4 x i32> %vec.ind)
call <4 x i32> @llvm.ctpop.v4i32(<4 x i32> %step.add)

\end{lstlisting}
\caption{LLVM IR program from Figure~\ref{fig:mot-llmir} after application of LLVM's existing vectorization pass.}
\label{fig:mot-vector}
\end{subfigure}
\label{fig:motivating}
\caption{Motivating Example}
\end{figure}

This section motivates the power of compiler-LLM cooperation in optimizing code with an example function. Throughout this paper, we provide examples using LLVM's \texttt{clang} compiler.

\sumpara{Input Function}
Figure~\ref{fig:mot-f} shows the source code of a function $f$ that that loops through values in range $[1,k]$. For each value, $f$ counts the number of bits that are set, and then sums the results; in short, $f$ returns the sum of ``population counts'' over a given range. $f$ is initially $O(k \log k)$.

\sumpara{Source Level LLM Optimization}
LLMs show potential for optimizing code on their own. Figure~\ref{fig:mot-llm} shows a version of $f$ optimized by Claude 3.7 Sonnet. Here, the LLM has introduced a single mathematical function that performs the bit counting for each value in the outer loop, leaving only the inner loop. $f$ is now $O(\log k)$.

\sumpara{LLM-based Optimization}
The potential of LLM-based code optimization is not limited to the source code level. Figure~\ref{fig:mot-ir} shows the LLVM IR code for the inner loop of function $f$, with two basic blocks; the rest of the program is omitted for brevity. An LLM is also capable of modifying the program at IR level, as shown in Figure~\ref{fig:mot-llmir}. Here, Claude 3.7 Sonnet is used again and identifies that the operation performed in the inner loop is bit counting. It therefore produces IR code that calls LLVM's bit-counting intrinsic \texttt{@llvm.ctpop.i32}, eliminating the inner loop and achieving $O(k)$ asymptotic runtime.\footnote{$O(k)$ runtime is achieved only if the underlying hardware supports bit counting in constant time; for instance, the x86-64 ISA provides \texttt{POPCNT} for 16-, 32-, and 64-bit integers.}

\sumpara{Compiler-LLM Cooperation}
Figure~\ref{fig:mot-vector} shows the results of applying existing compiler optimizations to the LLM-generated code in Figure~\ref{fig:mot-llmir}; LLVM is able to vectorize the calls to the population count intrinsic, placing $8$ repetitions ($2$ instructions with $4$ operations each) of the call within each iteration of the outer loop. For a processor with the right hardware, this further reduces the running time by a factor of $4\times$.

The existing compiler does not introduce the \texttt{ctpop} intrinsic on its own; only the LLM does so. On the other hand, the compiler is able to correctly vectorize the calls to the intrinsic while the LLM does not take advantage of this optimization opportunity. In this way, Figure~\ref{fig:mot-vector} shows the power of compiler-LLM cooperation: the LLM first spots an optimization that the compiler does not perform, and the compiler can subsequently perform a further optimization (vectorization) for which an LLM is ill-suited.

\section{Approach}
\label{sec:approach}
This section formally states the problem of compiler-LLM cooperation, and describes how we design it solution, including an LLM-based guiding agent, three level-specific optimization agents, and a testing agent.

\subsection{Preliminaries}
The problem of compiler-LLM cooperation is one of coordinating the lowering and rewriting tools provided by both compilers and LLMs. We here provide a formal statement of the problem, and in the rest of this section, describe our approach to realize compiler-LLM cooperation.

Assume we have a fixed ordered set of languages $\mathbb{L} = \{L_1, L_2, \dots, L_n\}$, where the ordering represents levels of abstraction; for example, we may have $L_1$ as C source code, $L_2$ as compiler IR, etc. A program $P$ written in language $L_i$ might be transformed in one of two ways.
\begin{definition}[Rewrite]
\label{def:rewrite}
    For some $i \leq n$, a function $f: L_i \rightarrow L_i$ is a \textit{rewrite} at level $i$.
\end{definition}

\begin{definition}[Lowering]
\label{def:lowering}
    For some $i, j \leq n$ such that $i < j$, a function $f: L_i \rightarrow L_j$ is a \textit{lowering} from level $i$ to level $j$.
\end{definition}
Rewrites and lowerings are the atomic operations that move between and within levels of abstraction in a compilation pipeline. A compiler then consists of a set of rewrites and lowerings:
\begin{definition}[Compiler]\label{def:compiler}
A compiler is a pair $(F,S)$ with a finite set of functions $F = \{f_1, f_2, \dots, f_k\}$ and a finite sequence $S \subset \mathcal{P}(F)$ such that
\begin{itemize}
    \item For all $f \in F$, $f$ is either a rewrite or a lowering,
    \item The first function of $S$, $S_1$ has domain $L_1$,
    \item The last function of $S$, $S_f$, has range $L_n$, and
    \item For each sequential pair $S_k, S_{k+1}$ in $S$, if the range of $S_k$ is a language $L_i$, then the domain of $S_{k+1}$ is also $L_i$.
\end{itemize}
\end{definition}

The properties of the sequence $S$ ensure that the compiler has a path involving lowerings and rewrites that stretches from the highest level of abstraction to the lowest. In practice, $S$ typically involves passing through every level of abstraction (no skipped levels). The existing LLVM compiler pipeline uses three levels of abstraction: C source code, LLVM IR, and x86 assembly. This compiler has $\left|F\right| = 3$:
\begin{itemize}
    \item The frontend is a lowering from C source code to LLVM IR;
    \item The ``middle-end'' is a rewriting at LLVM IR level, including all IR optimization passes; and
    \item The backend is a lowering from LLVM IR to assembly, which also includes some backend optimizations.
\end{itemize}
LLVM and other compilers also perform some light source code pre-processing (resolving \texttt{\#define} directives, for instance); we treat these as a constituent part of the frontend.

With the compiler in hand, we introduce level-specific optimization agents.
\begin{definition}[Level-Specific Agent]\label{def:level-agent}
    A level-specific agent is any algorithm that implements a rewrite function $f: L_i \rightarrow L_i$.
\end{definition}
In practice, we implement each level-specific agent with several LLM calls; Section~\ref{sec:level-specific} gives the details and variants of our level-specific agents. Last, we introduce testing agents:
\begin{definition}[Testing Agent]
    For a given level of abstraction $i$, a testing agent is an algorithm that implements a function $T_i: L_i \rightarrow \mathbb{R} \times \mathbb{R}$ such that $T_i(p)$ produces a pair $\left(T_\text{correct}, T_\text{perf}\right)$, where $T_\text{correct} = 1$ if the program is correct and less than $1$ otherwise, and $T_\text{perf}$ represents the performance of program $p$.
\end{definition}
In this work, we treat correctness as a proportion of tests passed and performance as a ratio of original runtime to final runtime. The problem of compiler-LLM cooperation is a problem of orchestrating rewrites provided by both LLMs and the compiler, lowerings provided by the compiler, and calls to the testing agent in order to achieve a correct program with good performance.
\begin{framed}
\centering
\textbf{Problem Statement}

Given a compiler $(F,S)$, a set of level-specific agents $A = \{a_1, a_2, \dots a_i\}$, a testing agent $T$, and an input program $p \in L_1$, construct a composition of the functions of $F$ and $A$, $\mathcal{C}$, such that 
\begin{itemize}
    \item $\mathcal{C}(p) \in L_n$, \ie, $(F \cup A, \mathcal{C})$ is a ``compiler'' under Definition~\ref{def:compiler}, and
    \item $T_\text{perf}\left(\mathcal{C}(p)\right)$ is maximized subject to the constraint $T_\text{correct}\left(\mathcal{C}(p)\right) = 1$
\end{itemize}
\end{framed}

The set of possible compositions of compiler and LLM-based components is infinite; moreover, each level-specific LLM-based generation is in general nondeterministic. It is therefore impractical to explore the entire search space, and we design an LLM agent to approximate a solution to the optimization problem. In the next section, we describe the design of our multi-agent system.

\subsection{Architecture}
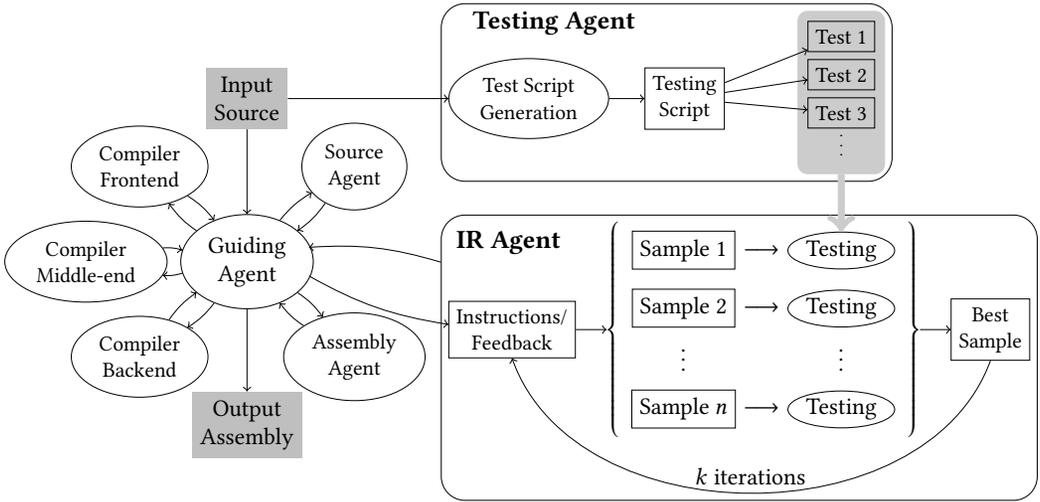
\begin{figure}
    \centering
    \resizebox{\linewidth}{!}{
    \begin{tikzpicture}[node distance=2cm]
%\begin{pgflowlevelscope}{\pgftransformscale{1}}
\begin{scope}[scale=1, shift={(0,0)}]
    \node[draw, ellipse, align=center, font = {\small}] (sourceAgent) at (1.6,1.4) {Source \\ Agent};
   
    \node[draw, ellipse, align=center, font = {\small}] (assemblyAgent) at (1.6,-1.4) {Assembly \\ Agent};
    
    \node[rectangle, align=center, fill=gray!50] (source) at (0,2.4) {Input \\ Source};

    \node[rectangle, align=center, fill=gray!50] (assembly) at (0,-2.4) {Output \\ Assembly};

    \node[draw, ellipse, align=center, font = {\small}] (frontend) at (-1.6,1.4) {Compiler \\ Frontend};

    \node[draw, ellipse, align=center, font = {\small}] (middleend) at (-2.4,0) {Compiler \\ Middle-end};

    \node[draw, ellipse, align=center, font = {\small}] (backend) at (-1.6,-1.4) {Compiler \\ Backend};

    \node[draw, ellipse, align=center] (planner) at (0,0) {Guiding \\ Agent};

    \draw[->] (planner) to [bend left=10] (sourceAgent);
    \draw[->] (sourceAgent) to [bend left=10] (planner);

    \draw[->] (planner) to [bend left=10] (assemblyAgent);
    \draw[->] (assemblyAgent) to [bend left=10] (planner);

    \draw[->] (planner) to [bend left=10] (frontend);
    \draw[->] (frontend) to [bend left=10] (planner);

    \draw[->] (planner) to [bend left=10] (middleend);
    \draw[->] (middleend) to [bend left=10] (planner);

    \draw[->] (planner) to [bend left=10] (backend);
    \draw[->] (backend) to [bend left=10] (planner);

    \draw[->] (source) -- (planner);
    \draw[->] (planner) -- (assembly);
\end{scope}

%\draw[->] (planner) to [bend left=10] (instructions);

\begin{scope}[scale=1,shift={(3,-1)}, anchor=west]
    \node[draw, rectangle, align=center, font= {\small}] (instructions) at (0,0) {Instructions/ \\ Feedback};

    \matrix [matrix of nodes, row sep=8, left delimiter=\lbrace,right delimiter=\rbrace, right=20pt of instructions] (mat)
      {
        \node[draw, rectangle] (sample1) {Sample 1}; & \node[] {$\longrightarrow$}; & \node[draw, ellipse, inner sep=1pt] (test1) {Testing}; \\
        \node[draw, rectangle] (sample2) {Sample 2}; & \node[] {$\longrightarrow$}; & \node[draw, ellipse, inner sep=1pt] (test2) {Testing}; \\
        \node[inner sep=-5pt] {\vdots}; & & \node[inner sep=-5pt] {\vdots}; \\
        \node[draw, rectangle] (sampleN) {Sample $n$}; & \node[] {$\longrightarrow$}; & \node[draw, ellipse, inner sep=1pt] (testn) {Testing}; \\
      };

    \node[draw, rectangle, align=center, font= {\small}, right=20pt of mat] (best) {Best \\ Sample}; 

    \draw[->] (instructions) to ($(mat.west) + (-0.7em, 0)$);

    \draw[->] ($(mat.east) + (0.7em, 0)$) to (best);

    \draw[->] (best.south) to [bend left=70] node[midway,above] (loop) {$k$ iterations} (instructions.south) ;

    \node[fit=(instructions)(best)(mat)(loop), inner sep=3pt, rounded corners=10pt, draw, label={[align=left, font={\large}, anchor=north west, xshift=3pt, yshift=-3pt]north west:\textbf{IR Agent}}] (irAgent) {}; 

\end{scope}

\begin{scope}[scale=1,shift={(3,2.4)}, anchor=west]

    \node[draw, ellipse, align=center, font= {\small}] (sg) at (0,0) {Test Script \\ Generation};

    \node[draw, rectangle, align=center, font= {\small}, above=of testn, right=15pt of sg] (script) {Testing \\ Script};

    \node[draw, rectangle, align=center, font= {\small}, yshift=26pt, anchor=center] (t1) at (test1 |- script) {Test 1};
    \node[draw, rectangle, align=center, font= {\small}, yshift=10pt, anchor=center] (t2) at (test1 |- script) {Test 2};
    \node[draw, rectangle, align=center, font= {\small}, yshift=-6pt, anchor=center] (t3) at (test1 |- script) {Test 3};
    \node[align=center, font= {\tiny}, below=4pt of t3, anchor=center] (td) {\vdots};

    \node[fit=(t1)(td), rounded corners=4pt, inner sep=4pt, draw=black!20, fill=black!20] (highlight) {};

    \node[draw, rectangle, align=center, font= {\small}, yshift=26pt, anchor=center] (t1) at (test1 |- script) {Test 1};
    \node[draw, rectangle, align=center, font= {\small}, yshift=10pt, anchor=center] (t2) at (test1 |- script) {Test 2};
    \node[draw, rectangle, align=center, font= {\small}, yshift=-6pt, anchor=center] (t3) at (test1 |- script) {Test 3};
    \node[align=center, font= {\tiny}, below=4pt of t3, anchor=center] (td) {\vdots};

    \draw[->] (sg) to (script);
    \draw[->] (script) to (t1);
    \draw[->] (script) to (t2);
    \draw[->] (script) to (t3);

    \node[fit=(sg)(t1)(td)(highlight)(highlight), rounded corners=10pt, inner sep=3pt, draw, label={[align=left, font={\large}, anchor=north west, xshift=10pt, yshift=0pt]north west:\textbf{Testing Agent}}] (agents) {}; 
    
\end{scope}

\draw[->] ($(planner.east) + (-2pt, -6pt)$) to [bend right=10] (instructions);
\draw[->] ($(irAgent.west) + (0pt, 40pt)$) to [bend right=10] ($(planner.east) + (-2pt, 6pt)$);

\draw[->] (source) to (sg);

\draw[->,draw=black!20,line width=3pt] (highlight) to (test1);
%\end{pgflowlevelscope}

\end{tikzpicture}}
    \caption{Our multi-agent architecture, showing the guiding agent, testing agent, and level-specific optimization agents. The source and assembly agents follow the same design as the expanded IR agent.}
    \label{fig:architecture}
\end{figure}

In order to orchestrate the lowerings and rewrites performed by both existing compiler components and LLM-based level-specific agents, we develop the multi-agent system shown in Figure~\ref{fig:architecture}---the organizational structure is similar to the planner-executor model of \citet{planAct}. The guiding agent, also an LLM, is empowered to make tool calls~\cite{tool-use} to components of the existing compiler and to level-specific optimization agents at each level of abstraction. When called, a level-specific agent performs a rewrite of the input program at its particular level of abstraction, and returns the optimized program to the guiding agent for further calls. In the next subsections, we describe each of these agents, what context they have access to, and how they communicate with each other.

\subsection{Guiding Agent}
The purpose of the guiding agent is to coordinate the actions of each level-specific agent and to interleave calls to these agents with calls to existing compiler components. We therefore design the guiding agent to have access to six ``tools''~\cite{tool-use} or function calls: three corresponding to the existing components of the LLVM compiler, and three level-specific LLM-based optimization agents. Once started, the guiding agent proceeds in a standard tool calling loop. First, it makes an inference call that determines, based on current context, which of the available tools to call. It then forms a call to the selected tool in appropriate syntax (in practice, a JSON string), and calls the selected tool. Once the tool has finished running, it returns, providing a string that gives feedback to the guiding agent. As we describe in the next section, this feedback from the level-specific agents includes performance and correctness measurements for the generated code. The feedback string is appended to the guiding agent's context, and the LLM is called again to select the next tool to call. Within the parameters of the provided tools, our design philosophy is to give the guiding agent maximum freedom to chose among the available tools.

There is one variable parameter which constrains the guiding agent, $b$, representing its compute budget. During a run of \tool, the guiding agent's context includes a budget tracker, which is updated after each tool call. We define a call to any level-specific agent to have unit cost, while calls to existing compiler components have cost $0$. Compiler components are free to use since, unlike level-specific agents which make multiple LLM inference calls, existing compiler components run locally on a CPU and have comparably small runtime. We thus enforce no limit on the number of calls to existing compiler components; however, the guiding agent's prompt directs it to make calls to level-specific optimization agents, which in practice prevents runaway calls to the compiler. The guiding agent's prompt also describes the budget limitation, allowing the LLM to reason about how to divide the budget most effectively among level-specific optimization agents. We enforce no architectural limits on the order of calls to level-specific agents, allowing repeated calls and backtracking. Indeed, the guiding agent may make repeated calls to a particular level-specific agent if it is unsatisfied with the results achieved---we evaluate the choices it makes in Section~\ref{sec:workings}.

Once the budget is exhausted, \tool stops calls to the guiding agent LLM and reports the end results to the user. If assembly level has not been reached (\textit{i.e.}, the run of \tool has not fulfilled Definition~\ref{def:compiler}), then the best-performing source or IR program produced so far is compiled with the existing compiler components and returned to the user as output. Intermediate optimization results from each call to a level-specific agent are also logged for later analysis.

\subsection{Level-Specific Agents}\label{sec:level-specific}

The responsibility of a level-specific optimization agent is to take as input code at a particular level of abstraction and output new code at that same level of abstraction which is correct and performant. Each level-specific agent is provided as context the input program and a string given by the guiding agent describing possible optimizations to attempt. After a run, it returns the optimized code along with a performance measurement to the guiding agent. In practice, \tool includes three level-specific agents: one each for C source code, LLVM IR, and x86 assembly. However, the design of the level-specific agents is not particularized to any one level of abstraction; we design a general level-specific agent and instantiate it once for each level, varying prompts and context only be replacing the name and description of the given level of abstraction. The design allows for easy generalization of \tool in case additional or different levels of abstraction are required.

A level-specific optimization agent could in principle simply call an LLM with its provided context; however, we include two classical enhancements for level-specific agents~\cite{scaling}: parallel sampling and iterative refinement feedback loops~\cite{selfRefine}. To that end, each level-specific agent is parameterized by two values:
\begin{itemize}
    \item $k$, the number of feedback loop iterations performed during a run of the agent, and
    \item $n$, the number of generations collected during each pass through the feedback loop.
\end{itemize}
A level-specific agent then works as follows. First, it makes $n$ calls to the LLM with identical context, producing a sample with $n$ optimized program variants. It then calls the testing agent on each of these generated programs, yielding both performance and correctness results. The best-performing correct generated program is selected and replaces the agent's input program, and the loop is repeated $k$ times, selecting the best-performing generation each time. We chose to perform sampling within each loop iteration rather than iterative refinement on each sample to limit computational resource use in practice. If a sample contains no correct programs, an aggregate of the feedback from the testing agent on the programs in the sample is appended to the level-specific agent's context and used to guide future optimization. The feedback includes compiler error messages, in the case that the generated program cannot be compiled to run, or a list of failing test cases, when the programs can be run but do not produce the correct results. After $k$ iterations of iterative refinement, calls to the level-specific agent's LLM are stopped and the best-performing program so far is returned to the guiding agent along with the measured performance. If no successful programs were generated, correctness feedback is returned to the guiding agent instead. 

We construct the level-specific agents to be modular and compatible with a large number of performance improvement strategies. For instance, as we discuss in Section~\ref{sec:related}, the level-specific agents could be improved with RL, ICL, or other enhancements as proposed by \citet{assembly}. However, the extent of the practical benefits of these techniques is largely orthogonal to our contribution; our framework supports any choice of model, learning, and reinforcement strategies for level-specific agents. We practically implement two basic enhancements, but leave to future work the question of which enhancements and configurations are best suited to each class of inputs.

\subsection{Testing Agent}\label{sec:testing}
In addition to the compiler and level-specific LLM-based optimization tools available to the guiding agent, we also develop a test generation agent whose responsibility is to generate test cases and run input programs to measure both performance and correctness. Design of the testing agent faces two important challenges. First, to adequately test performance, it must run an input program on test cases which are sufficiently numerous and diverse to exercise all possible behaviors of the program. Second, to fully measure performance, it must use test cases with a wide enough range to demonstrate the behavior of the program. We address both problems by constructing the testing agent to generate not individual test inputs, but rather a script which generates inputs in a deterministic manner.

We therefore split the testing agent into two phases, one that runs based on static information at the start of a run of \tool, and another which runs each time the testing agent is invoked by a level-specific agent, as shown in Figure~\ref{fig:architecture}. At the start of a run of \tool, the testing agent takes as context the initial input program and generates a script that can produce individual test inputs on demand. We divide the inputs produced by the testing script into two types: \textit{correctness exploration inputs} which are designed to probe the correctness of a generated program and, for instance, test edge cases; and \textit{large scale inputs}, which are designed to span a large range of input sizes to properly evaluate the runtime of the program under test. We define the test script generation task so that the testing script takes as input a parameter $i$ which determines the size of the large scale inputs. For each call to the testing agent, we generate $C$ correctness exploration inputs, and a further $L$ large scale inputs spanning multiple orders of magnitude.

In the second phase, which occurs every time a level-specific agent calls the testing agent, the script is run to generate $C+L$ inputs to the provided program; in our experiments, we have $C=10$ and $L=5$. The testing agent compiles the program into an executable from whatever level of abstraction was provided, and runs the program on each of the inputs, measuring the running time and comparing the program's output to the output from the reference initial input program, measuring both correctness and performance across all of the correctness exploration and large scale inputs. It then reports the performance and correctness results to the level-specific agent, including any compiler error messages encountered during the test.

Because different input programs have different purposes and take different categories of input (integers, floating-point numbers, strings, grids of data, \textit{etc}.), the contents of the test-generation script are highly program-dependent. The correctness exploration inputs are chosen during test script generation to explore as many possible program behaviors as possible. The large scale inputs, on the other hand, are designed to have size in proportion to the input parameter $i$---if $i$ increases by an order of magnitude, then the size of the input to the program under test should also increase by an order of magnitude. However, the meaning of input size differs for each program. For example, the testing script for a program which calculates the square of a number could just return $i$. Another program, like that in Example~\ref{ex:grid}, might expect a grid as input, and the test script could produce a random $i\times i$ grid. Still other programs could expect negative values as input, and have running time that grows as the input \textit{shrinks} in magnitude. We therefore rely on the LLM's reasoning ability to produce a script that generates inputs which grow as $i$ grows. Having inputs that can span orders of magnitude is critical to accurately measuring performance, as two programs with vastly different asymptotic performance may both appear to run quickly on a set of small inputs.

\section{Evaluation}
\label{sec:evaluation}

We evaluate our multi-level optimization method on a benchmark set of C programs, ablating to compare different configurations of \tool and analyzing optimization traces to help explain how multi-level optimization achieves speedups. We summarize our key results as follows:
\begin{framed}
\begin{itemize}
    \item Multi-level optimization with \tool outperforms optimization at any single level of abstraction, and a naive multi-level baseline.
    \item The guiding LLM regularly calls all level-specific agents and compiler components, and often repeats calls or backtracks to improve performance. Source-level optimization contributes the most to speedups.
    \item Performance improves as \tool is given a higher compute budget, and resources are better devoted to iterative refinement than parallel sampling.
\end{itemize}
\end{framed}

\subsection{Dataset}
We use as benchmarks C programs taken from the Project \codenet dataset~\cite{codenet}, which have been used to evaluate LLM-based optimization techniques in the past~\cite{assembly, learningEdits}. The dataset consists of 4,053 competitive programming \textit{problems}, with multiple \textit{programs} solving each one. We restrict our attention to C programs, for which 3,365 problems have at least one program; the total number of programs is 754,058. Many of these programs contain errors, do not compile, or are too small to provide meaningful performance measurements. A secondary contribution of our work is therefore a set of benchmarks useful for LLM-based performance improvement evaluation. We remove the approximately 30\% of programs in the dataset that do not compile in original form, as shown in Table~\ref{tab:filtering_programs}. We further exclude 205 problems, shown in Table~\ref{tab:filtering_problems}, for which our testing agent fails to produce testing scripts, equating to 15,969 programs. To fairly test our agent on programs where opportunities for optimization exist, we use variability across the solutions to a problem as a proxy for the existence of optimization opportunities~\cite{learningEdits}. For any problem with at least 10 programs, we sample from problems where the difference in running time between the slowest and fastest solutions is at least 10\%.
%We therefore sample programs work:
%In practice, we also require that all test cases finish within a 100 second timeout. 
This results in 564 problems, and we randomly sample 10 programs from each, giving a dataset of 5,640 programs with opportunities for optimization. We perform ablation studies and comparisons using a random sample of 100 programs to make experiments at scale tractable. Random sampling is in line with prior work using a 200-program sample~\cite{assembly}.

\begin{table}[]

\begin{subtable}[t]{\textwidth}
    \centering
    \begin{tabular}{|l|r|r|}
        \hline 
        \multicolumn{3}{|c|}{\textbf{Problems}} \\ \hline\hline
        Original Dataset & 3,365 & 100\% \\ \hline
        No Programs Compile & 199 & 5.9\% \\ \hline
        Test Generation Failed & 205 & 6.1\% \\ \hline
        Less Than 10 Programs & 1,064 & 31.6\% \\ \hline
        Low Running Time Variance & 1,333 & 39.6\% \\ \hline
        \rowcolor{gray!50} 
        Remaining Problems & 564 & 16.7\% \\ \hline
        
    \end{tabular}
    \caption{Breakdown of problems in the \codenet dataset. Each problem contains several solution programs.}
    \label{tab:filtering_problems}
\end{subtable}

\begin{subtable}[t]{\textwidth}
    \centering
    \begin{tabular}{|ll|r|r|}
\hline
\multicolumn{4}{|c|}{\textbf{Programs}} \\ \hline\hline
\multicolumn{2}{|l|}{Original Dataset} & 754,058 & 100\% \\ \hline
\multicolumn{2}{|l|}{Compiler Error} & 141,740 & 18.8\% \\ \hline
\multicolumn{2}{|l|}{Compiler Warning} & 64,775 & 8.6\% \\ \hline
\multicolumn{1}{|l|}{\multirow{2}{*}{Compiler Success}} & No Successful Test Cases & 15,969 & 2.1\% \\ \hhline{~|-|-|-}
\multicolumn{1}{|l|}{} & \cellcolor{gray!50} Remaining Programs & \cellcolor{gray!50} 547,543 & \cellcolor{gray!50} 70.5\% \\ \hline 
\end{tabular}
    \caption{Breakdown of program filtering across all problems in the \codenet dataset. We exclude the approximately 30\% of programs that do not compile in original form.}
    \label{tab:filtering_programs}
\end{subtable}
\caption{Dataset filtering by problems and programs.}
\end{table}

\subsection{Experimental Setup}
\tool is is implemented in the Strands Framework~\cite{strands} and made publicly available on Github\footnote{\url{https://github.com/amazon-science/acclaim}.}. It includes a central guiding agent that is given access to the core agents and \codein{clang} components as tools. We use LLVM version 18.1.3, and measure speedups against the performance of the original source code compiled with \texttt{clang -O3}. We perform experiments on a machine with 96 Intel Xeon Platinum 8275CL CPUs and 1TB RAM, accessing Claude 3.7 Sonnet via Amazon Bedrock. Each agent run is limited to one hour.

We design our implementation and experiments to investigate three research questions about compiler-LLM cooperation:
\begin{itemize}
    \item \textbf{RQ1: Effectiveness.} Does optimization at multiple levels outperform single-level LLM-based optimization?
    \item \textbf{RQ2: Process.} How does compiler-LLM cooperation achieve performance improvements, and what choices does the guiding agent make?
    \item \textbf{RQ3: Design.} What is the optimal configuration for multi-level LLM-based optimization?
\end{itemize}

\subsection{RQ1: Effectiveness}
The first research question asks whether multi-level optimization performs better than optimization at a single level of abstraction. To answer this question, we fix the total computational budget and perform experiments comparing  \tool against four baselines:
\begin{enumerate}
    \item\label{comparison-source} An agent restricted to performing source code optimizations only
    \item\label{comparison-ir} An agent restricted to performing IR optimizations only
    \item\label{comparison-assembly} An agent restricted to performing assembly optimizations only
    \item\label{comparison-portfolio} A per-input single-level portfolio of the above three approaches
\end{enumerate}

Comparisons~\ref{comparison-source}--\ref{comparison-assembly} investigate the performance of multi-level optimization relative to any particular choice of abstraction level. Comparison~\ref{comparison-assembly} is a proxy for comparison against the approach of \citet{assembly}, for which an open source implementation is not currently available.\footnote{Exact speedups, though, are not directly comparable with the approach of \citet{assembly} because of hardware differences and the lack of RL in our level-specific agent.} Comparison~\ref{comparison-assembly} takes the same fixed computation budget used for all the other comparisons, and divides it evenly among the source, IR, and assembly levels. It then performs level-specific optimization for each of these levels in parallel and at the end selects the best-performing optimization on a per-program basis. Comparison~\ref{comparison-source} is similar to source-only approaches like that of \citet{searchBased}. To our knowledge, no existing work has directly performed LLM-based optimization at the IR level in the CPU context; the closest analogue to Comparison~\ref{comparison-ir} is the approach of \citet{feedback}, which optimizes code size and predicts output LLVM IR.
%Comparison~\ref{comparison-portfolio} tests whether dynamic choice of abstraction level by our guiding agent can outperform a naive portfolio of level-specific optimizations. 

The results are shown in Table~\ref{tab:baselines}, which gives the geometric mean speedup for each configuration, along with the speedups at each percentile. \tool outperforms all of the baselines, though the source-only agent also performs well. For each case, Table~\ref{tab:baselines} shows that speedups are heavily clustered at the top of the distribution; speedups for the 75th and especially 99th percentile are far higher than the mean. The top-percentile speedups demonstrate that, at least for some programs, LLMs are capable of identifying optimizations at each level of abstraction.

\begin{figure}
    \centering
    \begin{subfigure}{0.45\textwidth}
        \includegraphics[width=\linewidth]{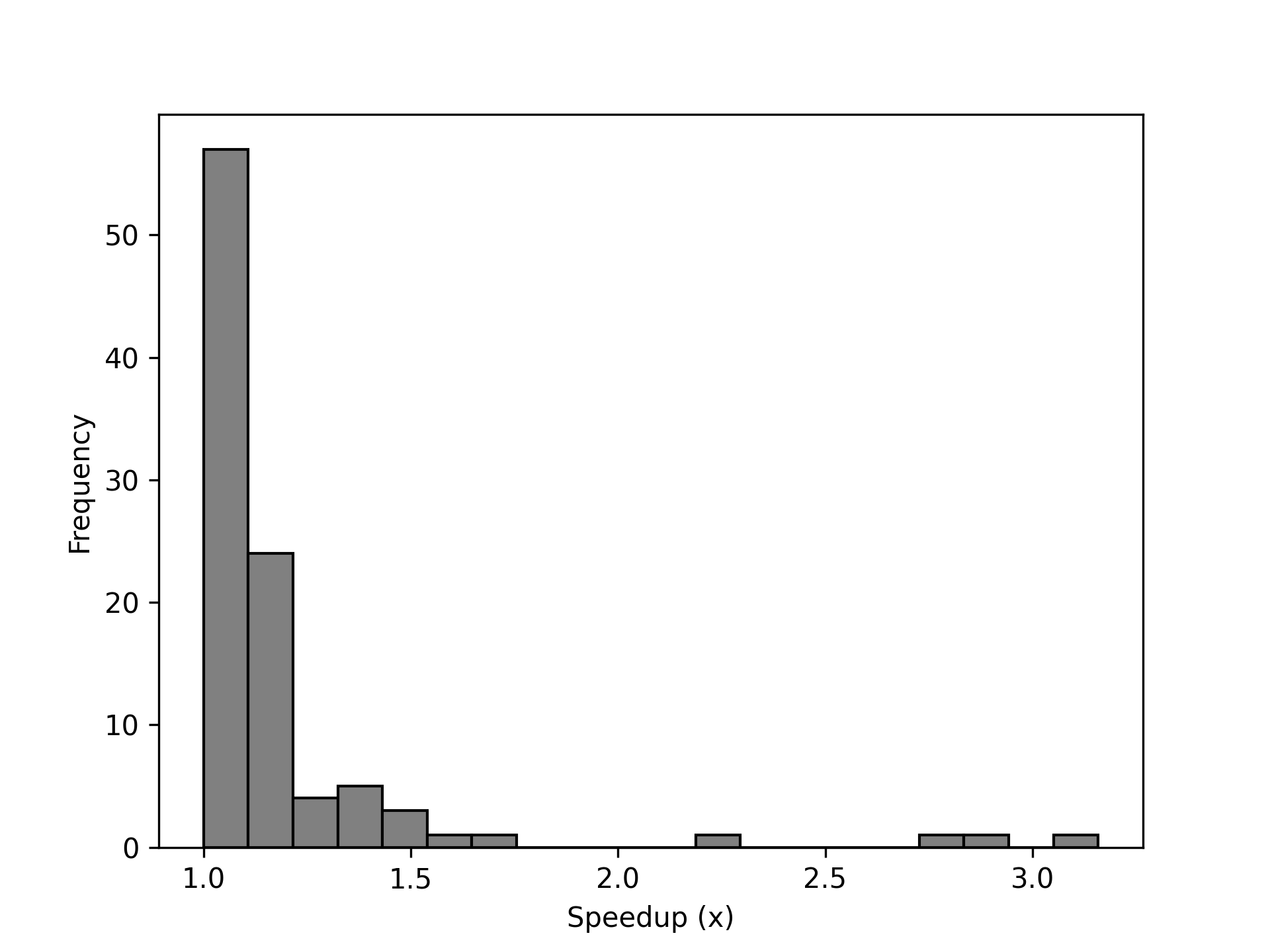}
        \caption{\toolst.}
        \label{fig:histo_tool}
    \end{subfigure}
    \hfill
    \begin{subfigure}{0.45\textwidth}
        \includegraphics[width=\linewidth]{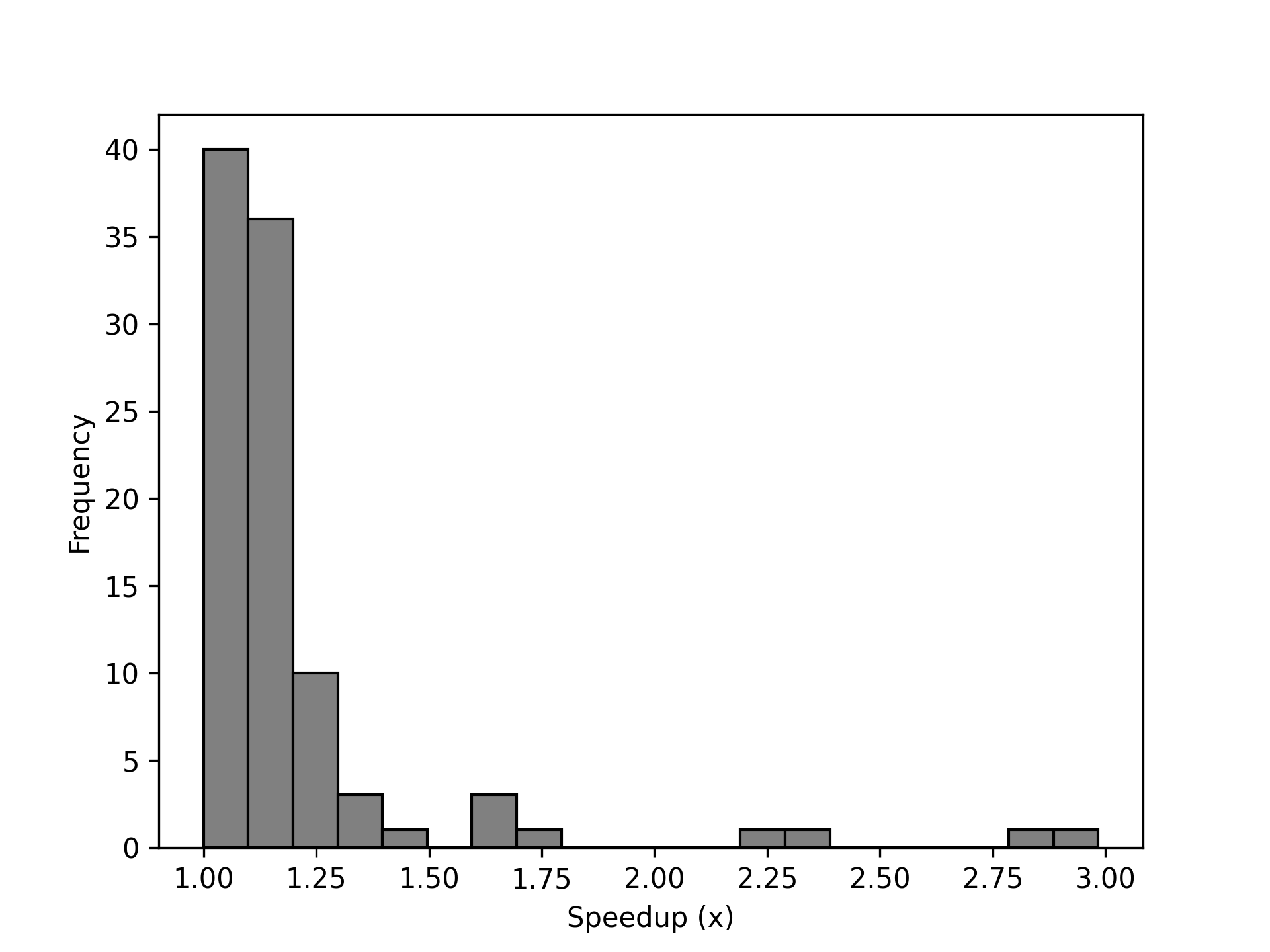}
        \caption{Source-only.}
        \label{fig:histo_source}
    \end{subfigure}

    \begin{subfigure}{0.45\textwidth}
    \includegraphics[width=\linewidth]{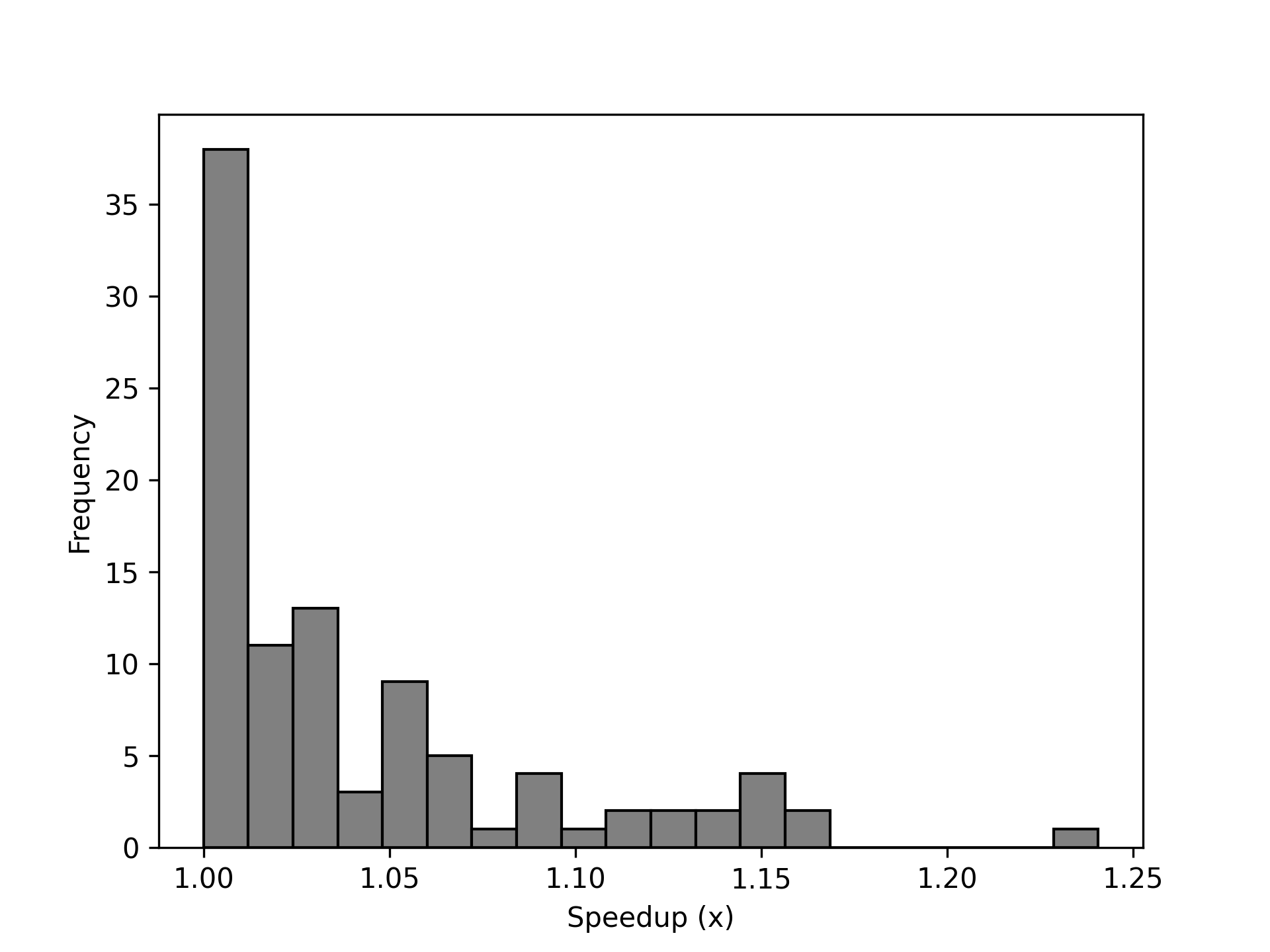}
    \caption{IR-only.}
    \label{fig:histo_ir}
    \end{subfigure}
    \hfill
    \begin{subfigure}{0.45\textwidth}
    \includegraphics[width=\linewidth]{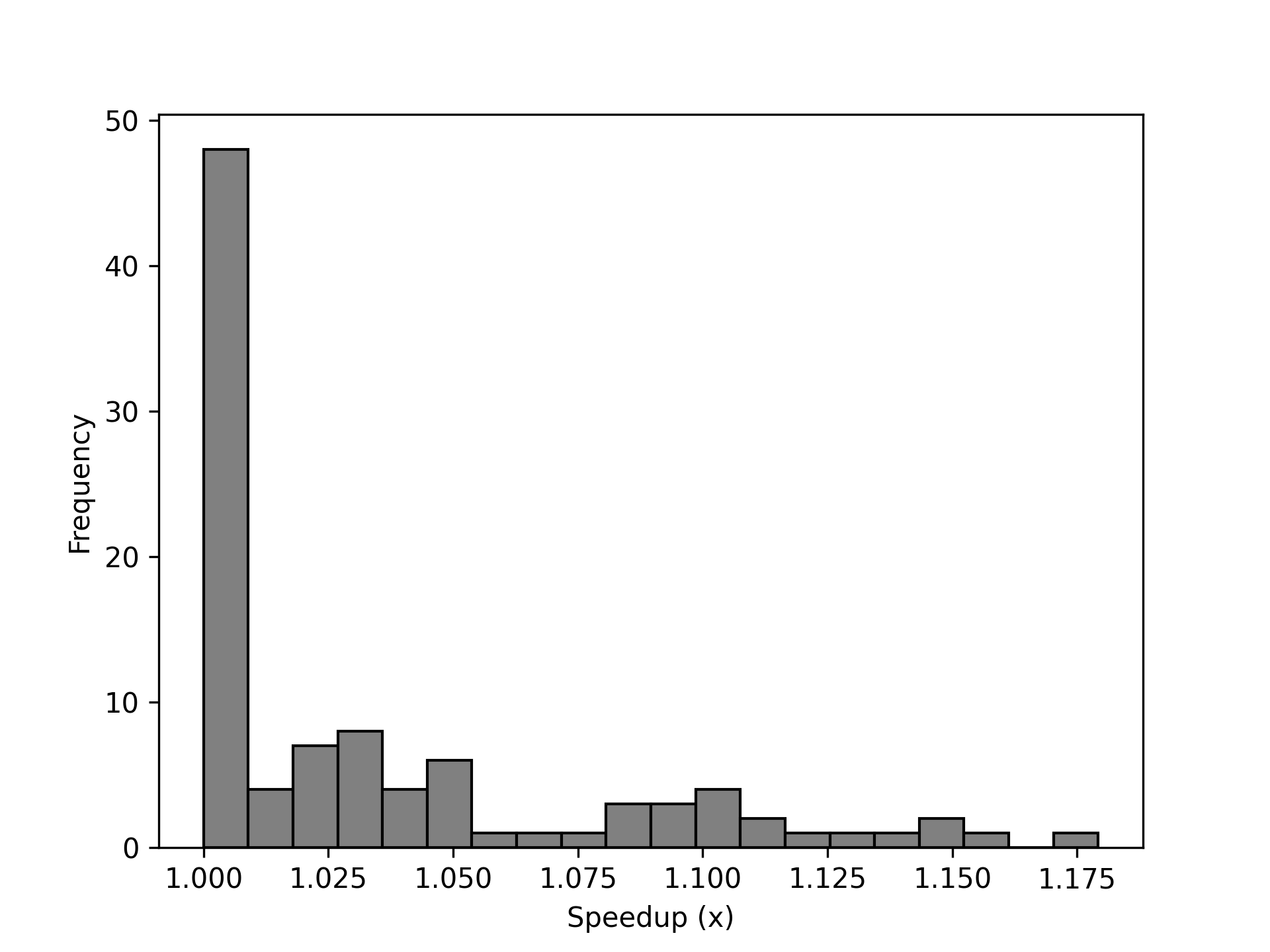}
    \caption{Assembly-only.}
    \label{fig:histo_assembly}
    \end{subfigure}
    \caption{Histogram of speedups produced by \tool and each level-specific baseline with $b=18, n = 2, k = 2$, scaled to show distribution of speedups. *Figure~\ref{fig:histo_tool} omits one program with speedup $1384\times$ for clarity; see the example.}
    \label{fig:histo}
\end{figure}

The results suggest that LLM-based optimization in general, and our multi-level approach in particular, achieves large speedups for some programs, while not improving others. Figure~\ref{fig:histo} shows the distribution of speedups for each configuration. LLM-based optimization at any level of abstraction does not substantially speed up most programs, with clusters of no or negligible speedup around $1.0\times$ in Figures~\ref{fig:histo_source}--\ref{fig:histo_assembly}, but a smaller number of programs sped up by much more---the absolute highest speedups, though, vary with each level, as shown by the differing $x$-axis scales. \tool follows the same trend, with very little improvement for most programs, but several programs are sped up by more than 50\%. The trend in the data arises because most programs do not contain many opportunities for optimization that existing compilers do not exploit. It is only rare programs where an LLM is able to exploit a high level optimization, which then yields a high speedup. The phenomenon of infrequent but large speedups during LLM-based optimization was also observed by \citet{assembly}, which found a much higher average speedup for the top percentiles of programs.

\begin{table}[]
    \centering
    \caption{Speedups ($\times$) achieved by our method and level-specific baselines.}
    \label{tab:baselines}
    \begin{tabular}{|l|c|c|c|c|c|}
    \hline
      \textbf{Configuration} & \textbf{Geo. Mean} & \textbf{25th} & \textbf{50th} & \textbf{75th} & \textbf{99th}\\ \hline
        Source-only (\ref{comparison-source}) & 1.18 & 1.06 & 1.12 & 1.18 & 2.98 \\ \hline
        IR-only (\ref{comparison-ir}) & 1.04 & 1.00 & 1.02 & 1.06 & 1.24 \\ \hline
        Assembly-only (\ref{comparison-assembly}) & 1.03 & 1.00 & 1.01 & 1.05 & 1.16 \\ \hline
        Single-level portfolio (\ref{comparison-portfolio}) & 1.02 & 1.00 & 1.00 & 1.02 & 1.20 \\ \hline
        \rowcolor{gray!50}
       \tool  & 1.25 & 1.04 & 1.10 & 1.16 & 16.97 \\ \hline
    \end{tabular}
    
\end{table}

\begin{example}\label{ex:grid}
The most striking example of high speedups for few programs is perhaps benchmark \texttt{p02644\_s879314742}; this program is a solution to a competitive programming problem involving path-finding on a square grid, where certain squares are blocked. The program reads as input the contents of the $\texttt{W} \times \texttt{H}$ grid, the maximum movement in a single step \texttt{K}, and the source and target coordinates. It then performs a breadth-first search through the grid to find the best path. \tool speeds up the program by $1384\times$ on overage over a set of 10 inputs with grid size between $5 \times 5$ and $1000 \times 1000$---the testing agent (Section~\ref{sec:testing}) is used to generate inputs spanning several orders of magnitude to fully assess the program's performance.

The performance improvements occur primarily at the source level, and span the program; we here highlight one important improvement to demonstrate the types of optimizations an LLM uses to outperform existing compilers. Figure~\ref{fig:malloc-old} shows a snippet of the program that allocates memory to hold the input grid; the program first allocates space for $H+2$ pointers, one for each row of the grid. It then separately allocates memory for each grid row in a loop, requesting memory from the operating system $H$ times, and then finally for the first and last rows which represent boundaries. \tool optimizes the program, producing the code shown in Figure~\ref{fig:malloc-new}. Rather than calling \texttt{malloc} in a loop, the optimized program allocates space for the entire grid, $(H+2) \times (W + 2)$ cells, in one go, asymptotically reducing the number of highly expensive system calls to allocate memory from linear in $H$ to constant. Adjustments to other parts of the program, including simplifications of the BFS implementation, also contribute to the overall speedup.

\begin{figure}
\begin{subfigure}{\textwidth}
\begin{lstlisting}[style=CodeStyle,escapechar=|]
c = (char**)|\colorbox{lightgray}{\codein{malloc(sizeof(char*) * (H + 2))}}|;
for (i = 1; i <= H; i++) {
    c[i] = (char*)|\colorbox{lightgray}{\codein{malloc(sizeof(char) * (W + 2))}}|;
    scanf("%s", &(c[i][1]));
}
c[0] = (char*)|\colorbox{lightgray}{\codein{malloc(sizeof(char) * (W + 2))}}|;
c[H+1] = (char*)|\colorbox{lightgray}{\codein{malloc(sizeof(char) * (W + 2))}}|;
\end{lstlisting}
\caption{Original program snipped that calls \texttt{malloc} $O(H)$ times.}
\label{fig:malloc-old}
\end{subfigure}
\begin{subfigure}{\textwidth}
\begin{lstlisting}[style=CodeStyle,escapechar=|]
char **grid = (char**)|\colorbox{lightgray}{\codein{malloc((H + 2) * sizeof(char*))}}|;
char *gridData = (char*)|\colorbox{lightgray}{\codein{malloc((H + 2) * (W + 2) * sizeof(char))}}|;
memset(gridData, '@', (H + 2) * (W + 2)); // Initialize all to walls

\end{lstlisting}
\caption{Program after source-level optimization by \tool, calling \texttt{malloc} only twice.}
\label{fig:malloc-new}
\end{subfigure}

\label{fig:malloc}
\caption{Source-level optimizations performed by \tool on benchmark \texttt{p02644\_s879314742}.}
\end{figure}

\end{example}

\subsection{RQ2: Process}\label{sec:workings}
For RQ2, we investigate how compiler-LLM cooperation works, analyzing the optimization logs it produces and quantifying the decisions made by the guiding agent.

\begin{table}[]
    \centering
    \caption{Speedup contributions, number of calls, proportion of individual samples correct, and proportion of samples that contain at least one correct solution. Shown for to each level-specific agent during a run of \tool, with $b = 18, n = 2, k = 2$. Speedup contributions do not add up to the overall speedup since they are multiplicative, rather than additive.}
    \label{tab:contributions}
    \begin{tabular}{|c|c|c|p{1.6cm}|p{1.8cm}|}
    \hline
    Level & Speedup Contribution & \# of Calls & \% Correct Generations & \% Correct Samples \\ \hline
     Source & $1.222\times$ & 427 & 53.5\% & 84.8\%\\ \hline
     IR & $1.012\times$ & 151 & 45.4\% & 41.0\% \\ \hline
     Assembly & $1.010 \times$ & 226 & 48.3\% & 57.4\% \\ \hline
     \rowcolor{gray!50}
     Overall & $1.248\times$ & 804 & 51.4\% & 69.4\% \\ \hline
    \end{tabular}
    
\end{table}

\sumpara{Level Contributions}
Table~\ref{tab:contributions} shows the frequency with which the guiding agent calls each level-specific agent and the speedup contribution of each. The speedup contributions show that source level optimization is the most effective, matching the results in Table~\ref{tab:baselines}, while IR and assembly optimization levels contribute far less to optimization. However, the contributions from each level are neither independent nor cleanly distinguishable. Interactions between different levels occur both as synergies in optimization (\ie, an optimization at one level enables or prevents an optimization at another level) and as budget-use choices made by the guiding agent (\ie, calling one level-specific agent because another was ineffective). These effects result in the difference difference between the level-specific \textit{contributions} shown in Table~\ref{tab:baselines} and the \textit{level-only baseline} results in Table~\ref{tab:baselines}.

Column ``\% Correct Generations'' of Table~\ref{tab:contributions} gives the proportion of individual LLM calls that pass all test cases---each of the $n$ members of the sample taken during level-specific optimization are counted independently. Column ``\% Correct Samples'' gives the percentage of level-specific agent samples \textit{in which at least one generated program was correct}---this is a ``pass@n'' metric. The proportion of correct generations is a reflection of the model's ability to produce correct code, while the proportion of correct samples measures the benefit of sampling $n$ times.

Since the guiding agent discards incorrect generations (\ie, they contribute no speedup), rates of incorrect generation are an important factor in defining each level's speedup contribution---and indeed, rates of correct generation are lower at IR and assembly level than for source code. However, the real difference lies in the number of level-specific agent runs where at least one of the $n$ samples is correct: this is $85\%$ for source but only $41\%$ for IR and $57\%$ for assembly. These results suggest that the IR and assembly agents are more likely to produce ``all-or-nothing'' samples, while at source code level, sampling produces greater variation.

\begin{figure}
    \centering
    \begin{tikzpicture}[node distance=2cm]
    \def\height{1.5}

    \node[draw, circle, align=center, text width=1.2cm, font = {\small}] (aa) at (2.5,\height) {Source};

    \node[draw, circle, align=center, text width=1.2cm, font = {\small}] (bb) at (7.5,\height) {Assembly};

    \node[fit=(aa)(bb), fill=black!10, rounded corners=8pt, label={[align=right, color=black!30, xshift=-5pt]left:\textbf{Level-Specific}\\\textbf{Agents}}] (agents) {};

    \node[draw, circle, align=center, text width=1.2cm, font = {\small}] (cc) at (2.5,-\height) {Frontend};

    \node[draw, circle, align=center, text width=1.2cm, font = {\small}] (dd) at (7.5,-\height) {Backend};

    \node[fit=(cc)(dd), fill=black!10, rounded corners=8pt, label={[align=right, color=black!30, xshift=-5pt]left:\textbf{Existing}\\\textbf{Compiler}}] (agents) {};

    \node[draw, circle, align=center, text width=1.2cm, font = {\small}] (entry) at (0,0) {Entry};

    \node[draw, circle, align=center, text width=1.2cm, font = {\small}] (source) at (2.5,\height) {Source};

    \node[draw, circle, align=center, text width=1.2cm, font = {\small}] (ir) at (5,\height) {IR};

    \node[draw, circle, align=center, text width=1.2cm, font = {\small}] (assembly) at (7.5,\height) {Assembly};

    \node[draw, circle, align=center, text width=1.2cm, font = {\small}] (frontend) at (2.5,-\height) {Frontend};

    \node[draw, circle, align=center, text width=1.2cm, font = {\small}] (middleend) at (5,-\height) {Middle-end};

    \node[draw, circle, align=center, text width=1.2cm, font = {\small}] (backend) at (7.5,-\height) {Backend};

    \node[draw, circle, align=center, text width=1.2cm, font = {\small}] (exit) at (10,0) {Exit};

    %\node[align=center, text width=1.2cm, font = {\tiny}] (toSource) at ($(backend) + (1.5, -1)$) {To Source};

    %\node[align=center, text width=1.2cm, font = {\tiny}] (fromBackend) at ($(backend) + (1.5, -1)$) {From Backend};

    \draw[dotted,->,line width=0.60mm] ($(backend) + (1, -0.5)$) -- ($(backend) + (1.5, -0.75)$);

    \draw[-,line width=0.60mm] (backend) -- ($(backend) + (1, -0.5)$) node[midway,right, yshift=3pt, font={\footnotesize}] {0.30};

    \draw[->,line width=0.60mm] ($(source) + (-1, 0.5)$) -- (source) node[midway,above=5pt, xshift=-9pt, font={\footnotesize}] {0.30};

    \draw[dotted,-,line width=0.60mm] ($(source) + (-1.5, 0.75)$) -- ($(source) + (-1, 0.5)$);

     \draw[->,line width=0.52mm] (entry) -- (source) node[midway,above=-3pt, xshift=-6pt, font={\footnotesize}] {0.26};

     \draw[->,line width=1.48mm] (entry) -- (frontend) node[midway,above=3pt, font={\footnotesize}] {0.74};
     
     \draw[->,line width=0.98mm] ($(source.south) + (0.3em, 0)$) -- ($(frontend.north) + (0.3em, 0)$) node[midway,auto, font={\footnotesize}] {0.49};

     \draw[->,line width=0.24mm] ($(frontend.north) - (0.3em, 0)$) -- ($(source.south) - (0.3em, 0)$) node[midway, auto, font={\footnotesize}] {0.12};

     \draw[->,line width=0.50mm] ($(ir.south) + (0.3em, 0)$) -- ($(middleend.north) + (0.3em, 0)$) node[midway,auto, font={\footnotesize}] {0.25};

     \draw[->,line width=0.22mm] ($(middleend.north) - (0.3em, 0)$) -- ($(ir.south) - (0.3em, 0)$) node[midway, auto, font={\footnotesize}] {0.11};

     \draw[->,line width=0.92mm] (backend) -- (assembly) node[midway,auto, font={\footnotesize}] {0.46};

     \draw[->,line width=0.62mm] (frontend) -- (ir) node[midway,above=6pt, font={\footnotesize}] {0.31};

     \draw[->,line width=0.92mm] (ir) -- (backend) node[midway,above=6pt, font={\footnotesize}] {0.46};

     \draw[->,line width=0.40mm] (frontend) -- (middleend) node[midway,auto, font={\footnotesize}] {0.20};

     \draw[->,line width=1.68mm] (middleend) -- (backend) node[midway,above, xshift=-5pt, font={\footnotesize}] {0.84};

     \draw[->,line width=0.22mm] (ir) -- (source) node[midway,auto, font={\footnotesize}] {0.11};

     \draw[->,line width=0.38mm] (assembly) -- (exit) node[midway,above,xshift=4pt, font={\footnotesize}] {0.19};

     \draw[->,line width=0.20mm] (backend) -- (exit) node[midway,above, xshift=-5pt, font={\footnotesize}] {0.10};

     \draw[->,line width=0.66mm] (frontend.south east) to [bend right=40] node[midway,above=-2pt, font={\footnotesize}] {0.33} (backend.south west);

     \draw[->,line width=0.22mm] ($(backend.south west) + (0.6em, -0.45em)$) to [bend left=40] node[midway,below=-2pt, font={\footnotesize}] {0.11} ($(frontend.south east) + (-0.6em, -0.45em)$);

     \draw[->,line width=0.12mm] ($(ir.north east) + (0em, 0em)$) to [bend left=60] node[midway,above left=2pt, font={\footnotesize}] {0.06} ($(exit.north) + (-0.5em, 0em)$);

     \draw[->,line width=0.18mm] ($(ir.north) + (0em, 0em)$) to [out=90, in=135, looseness=2.5] node[midway,right=6pt,yshift=3pt, font={\footnotesize}] {0.09} ($(ir.north west) + (0em, 0em)$);

     \draw[->,line width=0.68mm] ($(assembly.north west) + (0em, 0em)$) to [bend right=50] node[midway,above, xshift=22pt, yshift=-4pt, font={\footnotesize}] {0.34} ($(source.north east) + (0em, 0em)$);

     \draw[->,line width=0.12mm] ($(source.north) + (1em, -0.2em)$) to [bend left=60] node[midway,auto, font={\footnotesize}] {0.06} ($(exit.north) + (0.5em, 0em)$);

     \draw[->,line width=0.76mm] ($(assembly.south west) + (0em, 0em)$) to [out=-135, in=180, looseness=2.5] node[midway,left=-4pt,yshift=16pt, font={\footnotesize}] {0.38} ($(assembly.west) + (0em, 0.5em)$);

     \draw[->,line width=0.84mm] ($(source.north) + (0.3em, 0em)$) to [out=90, in=135, looseness=3.5] node[midway,above, font={\footnotesize}] {0.42} ($(source.north west) + (0.3em, 0.3em)$);

    % Style nodes as ellipses
    \tikzstyle{every node}=[ellipse, draw, minimum width=2cm, minimum height=1cm]
\end{tikzpicture}
    \caption{Graphical representation of \tool's transition frequencies as an automaton. Arrow width is proportional to frequency, and transitions with frequency less than 0.05 have been omitted for clarity. There is a single transition from the compiler backend to source-level optimization with frequency $0.30$ represented by disconnected dotted arrows.}
    \label{fig:markov}

\end{figure}
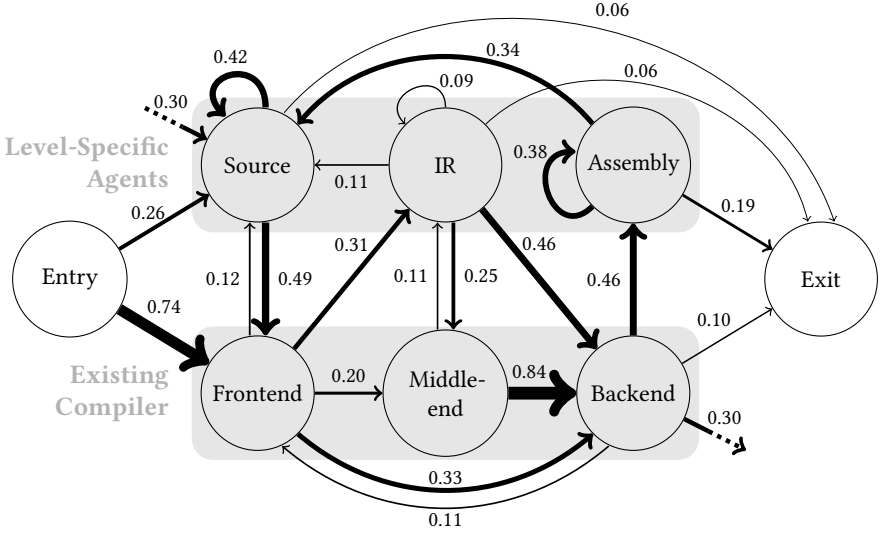
\sumpara{Agent Behavior}
We also investigate the choices made by the guiding agent during runs of \tool. Table~\ref{tab:contributions} shows that the guiding agent's calling behavior mirrors the level-specific agents' correctness: it calls source-level optimization at almost twice the rate of the other levels. In addition to the level-specific LLM-based optimization agents, the guiding agent also has access to the existing components of clang---the frontend, middle-end, and backend. Figure~\ref{fig:markov} shows the outbound transition frequency with which the agent transitions between each of the six tools it has access to. The most frequent transitions for the source and IR level agents are to the compiler frontend and backend, respectively; this shows the cooperation between LLM agents and the compiler enabled by \tool.

The initial transition frequencies (26\% for source-level optimization and 74\% for the compiler frontend) suggest that the guiding agent is not simply calling LLM-based source-level optimization, but instead producing intermediate results with which to generate feedback. The agent also frequently re-runs level specific agents---42\% of the time for source, 9\% for IR, and 38\% for assembly. This tendency may indicate that, when results produced by a level-specific tool are not satisfactory, the guiding agent calls the level-specific tool to ``try again''. One surprising trend is the frequency of transitions from the compiler backend to both the compiler frontend (11\% of the time) and source-level optimization (30\% of the time). This indicates an optimization workflow in which the guiding agent has produced a full optimization but then starts over from the highest level of abstraction.

\begin{example}
To illustrate how compiler-LLM optimization produces speedups across multiple levels, we show the example of benchmark \texttt{p00575\_s033386858}. This program models a game of consecutive days: for each day an action is performed, a player earns $a$ points, but $b$ points are earned after 7 consecutive days. The program takes as input a target number of points $c$ and computes the number of days needed to earn $c$ points. Figure~\ref{fig:trace} shows the performance of the input program at each intermediate point during a run of \tool; gray lines indicate unselected generations during level-specific agent sampling. Samples are not selected either because they have worse performance than that already achieved, or because they are not correct. Performance improvements are achieved at both the source and IR levels, with the primary contribution coming from an IR transformation.

\begin{figure}

\begin{subfigure}{\textwidth}
\begin{lstlisting}[style=CodeStyle,escapechar=|]
%10 = add nsw i32 %8, %9
%11 = sdiv i32 %6, %10
%12 = |\colorbox{lightgray}{\codein{mul nsw i32 \%11, 7}}|
%13 = srem i32 %6, %10
%14 = icmp slt i32 %13, %8
\end{lstlisting}
\caption{Initial LLVM IR program before optimization.}
\label{fig:evaluation-ir-old}
\end{subfigure}

\begin{subfigure}{\textwidth}
\begin{lstlisting}[style=CodeStyle,escapechar=|]
%10 = add nsw i32 %8, %9
; Combine division and multiplication operations
%11 = sdiv i32 %6, %10
%12 = |\colorbox{lightgray}{\codein{shl nsw i32 \%11, 3}}|
%13 = |\colorbox{lightgray}{\codein{sub nsw i32 \%12, \%11}}|
%14 = srem i32 %6, %10
%15 = icmp slt i32 %14, %8
\end{lstlisting}
\caption{Program after optimization by the IR level-specific agent. The comment marked with ``;'' was inserted by the agent.}
\label{fig:evaluation-ir-new}
\end{subfigure}

\caption{Exceprt of IR code during optimization of benchmark \texttt{p00575\_s033386858}.}
\label{fig:evaluation-ir}
\end{figure}

Figure~\ref{fig:evaluation-ir} shows an excerpt of the program during the second round of IR-level optimization. The IR agent converts a multiplication by the constant $7$ (Figure~\ref{fig:evaluation-ir-old}) into the two instruction sequence shift left by $3$, equivalent to multiplication by $8$, and then subtract $1$ (Figure~\ref{fig:evaluation-ir-new}). LLVM is capable of performing this optimization, but for the subject program does not do so, even at \texttt{O3}; the optimization is also not performed during the compiler backend. In this instance, LLVM's cost model, which is designed to be general, mispredicts the cost of the multiplication operation and chooses not to replace it with a shift instruction; the LLM-based IR agent does so instead, leading to a speedup increase from $1.06\times$ to $1.16\times$.

\pgfplotstableread[
    col sep=comma]{
index,start,end
0,1,0.918586163030541
0,1,0.880838546926923
1,1,0.980262927296613
1,1,0.951983457190144
2,1,0.997792445429163
3,1,0.923722023933854
3,1,0.890490461764877
4,1,0.972608796758133
5,1,1.02275750143007
5,1,1.2770058675257
6,1.02275750143007,0.968151678718604
6,1.02275750143007,0.941301924023127
7,1.02275750143007,0.983961095049215
7,1.02275750143007,1.06267005257485
8,1.06267005257485,1.16265724622712
9,1.16265724622712,1.06286406738316
9,1.16265724622712,1.12004532547107
10,1.16265724622712,1.09782000731468
10,1.16265724622712,1.12156084554875
11,1.16265724622712,0.953187961008277
11,1.16265724622712,0.95948507125497
}\graytable

\pgfplotstableread[
    col sep=comma]{
index,start,end
0,1,1
1,1,1
2,1,1
3,1,1
4,1,1
5,1,1.02275750143007
6,1.02275750143007,1.02275750143007
7,1.02275750143007,1.06267005257485
8,1.06267005257485,1.16265724622712
9,1.16265724622712,1.16265724622712
10,1.16265724622712,1.16265724622712
11,1.16265724622712,1.16265724622712
}\blacktable

\begin{figure}
\begin{tikzpicture}
    \begin{axis}[
        width=0.8\linewidth, 
        height=5cm,
        xlabel={Optimization Round $\longrightarrow$},
        ylabel={Speedup ($\times$)},
        xmin=0,
        xmax=12,
        ymin=0.7,
        ymax=1.3,
        xtick=\empty
        %legend pos=north west, % Position the legend
    ]

            % Read the file contents into a macro

    \pgfplotstablegetrowsof{\graytable} % gets number of rows
    \pgfmathsetmacro{\grayrows}{\pgfmathresult} % sets \nrows
    
    \pgfmathtruncatemacro{\lastgray}{\grayrows-1} % sets \lastrow to (number of rows - 1)

    \foreach \row in {0,...,\lastgray} {
        \pgfplotstablegetelem{\row}{index}\of\graytable
        \let\index=\pgfplotsretval
        \pgfplotstablegetelem{\row}{start}\of\graytable
        \let\mstart=\pgfplotsretval
        \pgfplotstablegetelem{\row}{end}\of\graytable
        \let\mend=\pgfplotsretval
        
        \pgfmathparse{\index+1}
        \let\nextindex=\pgfmathresult

        \expandafter\addplot+[mark=none,color=black!30,line width=2pt, solid] coordinates { (\index,\mstart) (\nextindex,\mend) };
        %\node[above] at (axis cs:9,1) {tt"\mcolor"};
    }

    \pgfplotstablegetrowsof{\blacktable} % gets number of rows
    \pgfmathsetmacro{\blackrows}{\pgfmathresult} % sets \nrows
    
    \pgfmathtruncatemacro{\lastblack}{\blackrows-1} % sets \lastrow to (number of rows - 1)

    \foreach \row in {0,...,\lastblack} {
        \pgfplotstablegetelem{\row}{index}\of\blacktable
        \let\index=\pgfplotsretval
        \pgfplotstablegetelem{\row}{start}\of\blacktable
        \let\mstart=\pgfplotsretval
        \pgfplotstablegetelem{\row}{end}\of\blacktable
        \let\mend=\pgfplotsretval
        
        \pgfmathparse{\index+1}
        \let\nextindex=\pgfmathresult

        \expandafter\addplot+[mark=none,color=black,line width=2pt, solid] coordinates { (\index,\mstart) (\nextindex,\mend) };
        %\node[above] at (axis cs:9,1) {tt"\mcolor"};
    }

    \addplot+[mark=none,color=black,dotted] coordinates { (2,0) (2,2) };
    \addplot+[mark=none,color=black,dotted] coordinates { (4,0) (4,2) };
    \addplot+[mark=none,color=black,dotted] coordinates { (6,0) (6,2) };
    \addplot+[mark=none,color=black,dotted] coordinates { (8,0) (8,2) };
    \addplot+[mark=none,color=black,dotted] coordinates { (10,0) (10,2) };

    \node[font={\small}] at (axis cs:1, 0.8) {Source};
    \node[font={\small}] at (axis cs:3, 0.8) {IR};\node[font={\small}] at (axis cs:5, 0.8) {Assembly};\node[font={\small}] at (axis cs:7, 0.8) {Source};
    \node[font={\small}] at (axis cs:9, 0.8) {IR};\node[font={\small}] at (axis cs:11, 0.8) {Assembly};

    \end{axis}
\end{tikzpicture}
\caption{Trace of optimization for benchmark \texttt{p00575\_s033386858}, with $b = 6$, $n = 2$, $k = 2$. Gray lines show samples within each level-specific loop, including incorrect samples; black lines show the selected best correct optimized version.}
    \label{fig:trace}
\end{figure}
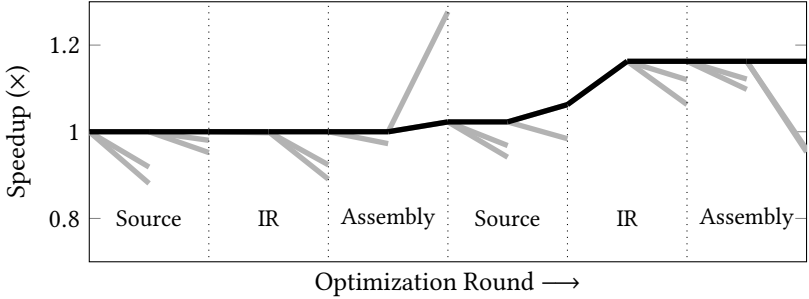
\end{example}

\subsection{RQ3: Design}\label{sec:ablations}
To investigate the best design of a compiler-LLM cooperation system, we perform several ablations to study the impact of changes in compute budget and configuration settings on the performance of \tool. We consider how compute budget should be distributed, how the performance of compiler-LLM cooperation depends on overall compute budget, and which LLMs are best-suited to the task.

\sumpara{Model Selection}
We evaluate compiler-LLM cooperation with six language models available as of August 2025, with the results shown in Table~\ref{tab:ablations}. As in Table~\ref{tab:contributions}, ``\% Correct Generations'' is the proportion of individual LLM calls that produce a correct program, while ``\% Correct Samples'' gives the number of samples with at least one correct generation. We evaluate with Claude 3.7 Sonnet, in addition to five open-source models: two variants of Llamma (\texttt{llama4-maverick-17b-instruct} and \texttt{llama3-3-70b-instruct}), Qwen 2.5, and two different sizes of Qwen 3 models (4b and 30b, with mixture-of-experts). The results show that Claude 3.7 Sonnet outperforms all of the open-source models. Most of the difference is attributable to differences in the number of correct generations---since each incorrect generation results in a sample being discarded, and thus no speedup, models that only rarely produce correct samples cannot achieve any speedup. By investigating the optimization traces, we find that one reason for the poor correctness performance of the Qwen and Llama models is related to their tool-calling ability rather than their ability to generate code; indeed, many calls to level-specific agents fail because the guiding agent is unable to generate well-formed tool calls and outputs plain text calls which are ineffective. Tool-calling is an integral part of the compiler-LLM cooperation framework, though it may be possible to add an additional tool-calling harness~\cite{nl-tools} or to improve models' compiler-related tool calling ability through further context engineering or finetuning with in-context examples~\cite{toolLLM}.

\sumpara{Distribution of Compute Budget}
The design of \tool includes three tunable parameters for computation budget: $b$, the number of level-specific calls the guiding agent can make, $n$, the number of parallel samples collected in each loop of a level-specific agent, and $k$, the number of feedback loop iterations performed by each level-specific agent. The computational cost of running experiments prevents a full scan of values of $b$, $n$, and $k$, so we compare three versions of \tool with fixed $b$ and cap the quantity $n \times k$; the results are shown in Table~\ref{tab:ablations}. We find that, using the most successful model (Claude 3.7 Sonnet), devoting more computational resources to feedback loops instead of parallel sampling ($n=1, k = 4$) produces the best results. An even division between sampling and feedback loops ($n=2, k=2$) is not far behind, while devoting all resources to parallel sampling instead of feedback loops performs worse ($1.11\times$ speedup). Iterative refinement with the feedback loop helps the level-specific agents reason about and correct their mistakes; in the example above, for instance, after receiving failed test results, the IR-specific agent proceeds to the next iterative refinement loop with feedback indicating what to do next: ``Consider using register hints (like register variables) and further simplify the conditional logic''.

The measurements of the number of correct generations and samples shed some light on the reason for the results: the proportion of correct generations is highest for $k=4$, suggesting that the level-specific agent's feedback loop increases the frequency with which the model produces correct code. While $n=4$ (a larger sample size) results in a larger spread between the number of correct generations and number of correct samples, the decrease in feedback loops reduces the absolute number of correct generations, and thereby the overall speedup.

\sumpara{Compute Time Scaling}
To test whether increasing overall computation budget improves model results, we triple the overall compute budget and perform experiments with the best-performing model and parameters $b = 18$, $n = 2$, $k = 2$; we keep $n$ and $k$ equal to eliminate any effects from changing the distribution of compute budget between $n$ and $k$. The results show that increasing the compute budget improves overall performance, increasing mean speedups from $1.19\times$ to $1.25\times$. This result suggests that the performance improvements achieved by compiler-LLM cooperation scale with increases in compute budget, and that this increase has not yet leveled off by $n=18$. Whether it is worthwhile to invest greater resources to achieve small performance improvements likely depends on the tradeoffs of a particular application domain. Since some compiled programs may be run billions of times by millions of users, even small incremental improvements to runtime could merit a high inference cost.

\begin{table}[!t]
    \centering
    \caption{Results for ablation studies of different configurations of \tool.}
    \label{tab:ablations}
    \begin{tabular}{|p{3cm}|c|c|c|c|c|c|}
    \hline
    Model & $b$ & $n$ & $k$ & Speedup  & \% Correct Generations & \% Correct Samples \\ \hline
    \rowcolor{gray!50}
    \texttt{claude-3.7-sonnet} & 18 & 2 & 2 & 1.25 & 51.4\% & 69.4\% \\ \hline
    \texttt{claude-3.7-sonnet} & 6 & 1 & 4 & 1.19 & 54.5\% & 64.4\% \\ \hline
    \texttt{claude-3.7-sonnet} & 6 & 2 & 2 & 1.16 & 48.0\% & 48.1\% \\ \hline
    \texttt{claude-3.7-sonnet} & 6 & 4 & 1 & 1.11 & 42.1\% & 48.4\% \\ \hline
    \texttt{llamma4-maverick-} \texttt{17b-instruct} & 6 & 2 & 2 & 1.01 & 13.0\% & 31.9\% \\ \hline
    \texttt{llama3-3-70b} \texttt{-instruct} & 6 & 2 & 2 & 1.00 & 0\% & 0\% \\ \hline
    \texttt{qwen2.5-coder-7b} & 6 & 2 & 2 & 1.00 & 0\% & 0\% \\ \hline
    \texttt{qwen3-4b} & 6 & 2 & 2 & 1.00 & 1.3\% & 16.8\% \\ \hline
    \texttt{qwen3-30b-a3b} & 6 & 2 & 2 & 1.04 & 8.3\% & 27.7\% \\ \hline
    \end{tabular}
    
\end{table}

\section{Discussion and Future Work}
\label{sec:discussion}

\subsection{Generalization to Other Compilers}\label{sec:discussion-compilers}
We realize \tool using the LLVM framework~\cite{llvm}; LLVM is the most commonly-used compiler in prior work~\cite{llmCompiler, feedback, alive2}. However, the structure of compiler-LLM cooperation is general, and could be applied to other compilers or computing environments. In the CPU context, GCC has the same frontend, middle-end, backend structure as clang and its own IR~\cite{gcc}; \tool could be adapted to cooperate with GCC with minimal engineering effort. Other source languages, like Rust, could also be targeted. Extend compiler-LLM cooperation to software targeting GPUs is also of great practical importance. As in the CPU context, the GPU software stack includes several levels of abstraction: source code may be written in PyTorch, then compiled to the Triton language~\cite{triton}. Triton kernels are then lowered to MLIR and then to LLVM IR where they benefit from the same LLVM optimization passes used for CPU programs. Finally, they are lowered to a low-level hardware-specific ISA like AMD GCN or Nvidia PTX. Existing work has already proposed level-specific rewrites for PyTorch~\cite{kernelbench, kevin} and Triton~\cite{kernelLLM}, so future work could integrate these level-specific optimization strategies with the existing components of the Triton and LLVM compilers to bring the benefits of compiler-LLM cooperation to GPU kernels as well---optimization of GPU kernels is of particular interest to improve the performance of LLMs themselves.

Compiler-LLM cooperation may also have applications in less closely-related applications. For instance, in Definitions~\ref{def:rewrite}--\ref{def:compiler}, we state lowerings and rewrites as the atomic operations of compilers, but recently, interest has also turned to inverse lowering, \textit{i.e.}, lifting~\cite{lifting}, including by using LLMs~\cite{llmDecomp}. Future work could address whether cooperation between existing compiler components and verification tools like \tool may be able to improve the correctness of LLM-based decompilations.

\subsection{Correctness and Verification}
Our use of testing to check the correctness of LLM-generated programs is in line with existing work in both the CPU~\cite{assembly, searchBased, learningEdits} and GPU~\cite{kernelbench, kevin} use cases. However, testing is not infallible, and may miss bugs or correctness issues in the LLM-generated code at any of the three levels of abstraction we address. We mitigate the risk of incorrect code generation through the design of the testing agent (\ref{sec:testing}), which can produce additional test inputs on demand and includes test cases designed to probe the correctness of the generated code. The structure of compiler-LLM cooperation itself helps to ensure correctness by combining the guaranteed correctness of existing compiler optimizations\footnote{Compilers perform correct transformations, modulo compiler bugs. We expect that compiler bugs affecting correctness are vanishingly rare relative to the frequency of incorrect LLM-generated code.} with the possibly error-bearing optimization potential of LLM-based code generation. The design of the testing agent also does not share full test cases with the level-specific optimization agents in order to reduce the likelihood of reward hacking~\cite{reward-hacking}; we have not qualitatively observed reward hacking behavior in the reasoning traces from runs of \tool.

Nevertheless, future work could address the problem of fully verifying the correctness of output programs from compiler-LLM cooperation. Such verification could take the form of translation validation~\cite{tv-necula}, fully checking the equivalence of the input and output programs. Existing tools like Alive2~\cite{alive2} may be useful for the IR portions of this task, while more sophisticated approaches would be required to verify the equivalence of input C and output assembly programs. Better translation validation methodologies would increase confidence in the programs produced by \tool, but their development is largely orthogonal to the strategy of compiler-LLM cooperation; the strategy is modular and would be compatible with the drop-in replacement of a testing agent with a verification agent instead. A verification agent which performs translation validation may even be able to provide more granular feedback to the LLM during runs of a level-specific agent.
\subsection{Computational Cost}
One important drawback of compiler-LLM cooperation is high computational cost: a run on a single program requires $b \times n \times k$ inference calls to a model, each with a number of tokens proportional to input program size. The results of Section~\ref{sec:ablations} show that increasing compute budget can increase the performance gains produced by \tool. Future work may determine where the point of diminishing return lies. This point likely depends heavily on input program features, the selected model, and the performance of the existing compiler components. The use case of the input program matters too: for a program that goes on to wide use as infrastructure software, devoting significant computing resources to achieve a small speedup may be practically worthwhile, while a program that will only be run one time by a single user may not merit the cost of LLM inference at all.

Future work could also explore methods of reducing the computational burden associated with compiler-LLM cooperation. The modular design of \tool allows mixing and matching of models---a larger, more expensive model could be used for the guiding agent, for instance, while a smaller, possibly finetuned, model is used for each level-specific agent. Computational burden could also be reduced by focusing LLM-based optimization only to parts of the input program; indeed, the guiding agent could take on the additional task of selecting portions of the input program which are most likely to be amenable to LLM-based optimization at the level of functions, basic blocks, or even individual instructions. This division may also boost correctness by decreasing the ``attack surface'' on which LLM-based code generation can introduce bugs.

\section{Related Work}
\label{sec:related}
\begin{table}[]
    \centering
    \caption{Summary of related work on LLM-based code optimization.}
    \label{tab:related}
    \small
    \begin{tabular}{|C{2cm}|C{1.3cm}|C{1.2cm}|C{0.8cm}|C{1cm}|C{1.3cm}|C{1.4cm}|c|}
        \hline
         & \multicolumn{3}{c|}{GPU} & \multicolumn{4}{c|}{CPU} \\ \hline
        Approach &\citet{kernelbench} & KernelLLM \cite{kernelLLM} & Kevin \cite{kevin} & SBLLM \cite{searchBased} & \citet{learningEdits} & SuperCoder \cite{assembly} & \cellcolor{gray!50} \tool \\ \hline
        Source Optimization & \cmark & & & \cmark & \cmark & & \cellcolor{gray!50}  \cmark \\ \hline
        Intermediate Optimization & & \cmark & & &  & & \cellcolor{gray!50}  \cmark \\ \hline
        Low-level Optimization & & & \cmark & & & \cmark & \cellcolor{gray!50}  \cmark \\ \hline
        Repeated Sampling & \cmark & & \cmark & \cmark & \cmark & & \cellcolor{gray!50}  \cmark \\ \hline
        Iterative Refinement & \cmark & & \cmark & \cmark & & & \cellcolor{gray!50}  \cmark \\ \hline
        Reinforcement Learning & & \cmark & \cmark & & & \cmark & \cellcolor{gray!50}  \\ \hline
        Supervised Fine-Tuning & & \cmark & & & \cmark & & \cellcolor{gray!50}  \\ \hline
    \end{tabular}

\end{table}

\sumpara{LLM-based Code Optimization}
Our approach is most closely related to the body of work on LLM-based code optimization. Existing approaches in this area have used LLMs to optimize source code; SBLLM~\cite{searchBased} integrates LLM-generated re-writes with an evolutionary search for better-performing code, while \citet{learningEdits} train an LLM with supervised fine-tuning and self-play to optimize C source code. SuperCoder~\cite{assembly} performs the same task at the level of x86 assembly; it employs reinforcement learning to train LLMs to generate fast and correct assembly code, given as input C source code and compiler-generated assembly. Constrained decoding~\cite{grammar-decoding, xgrammar} has been proposed as a method to boost the correctness of generated code by limiting the output of the LLM to match the grammar of the given language. LLMs have also been employed to optimize programs in the GPU context. Here, the seminal work is KernelBench~\cite{kernelbench}, which collects a dataset and establishes a baseline for LLM-driven generation of GPU kernels. It operates at the level of PyTorch, but permits the LLM to insert lower-level kernel code as well, and uses both repeated sampling and iterative refinement with feedback derived from compilation and testing of the generated kernels. Other work has improved upon the base of KernelBench~\cite{metrKernel, nvidiaKernel}. KernelLLM~\cite{kernelLLM} optimizes code in the Triton language, an intermediate between PyTorch and low-level CUDA kernels and takes advantage of both reinforcement learning and supervised fine-tuning. The state-of-the-art approach is Kevin~\cite{kevin}, which uses multi-turn reinforcement learning to further improve performance on the KernelBench dataset.

Table~\ref{tab:related} summarizes the relationship between these approaches. \tool operates on CPU-targeted programs and implements repeated sampling and iterative refinement with a feedback loop as do existing approaches. Our approach differs in that it integrates optimization at multiple levels of abstraction into a single optimization workflow, and includes a guiding agent to orchestrate the process. \tool provides a proof of concept realization of compiler-LLM cooperation, including the two most common performance improvements in the existing literature: repeated sampling and iterative refinement. We leave to future work the question of how much practical improvement could be gained by incorporating other more advanced improvements like reinforcement learning and supervised fine-tuning, noting only that the design of our level-specific agents would be compatible with both.

\sumpara{Integration of Compilers and LLMs}
In addition to direct LLM-based optimization of input programs, existing work has harnessed LLMs to direct existing compiler passes and settings. CompileAgent~\cite{compileAgent} gives existing compilers and other build tools as tools to an LLM agent, and constructs a feedback loop by passing compiler error messages back to the LLM. LLMCompiler~\cite{llmCompiler} performs model training to enhance understanding of programs at IR level, and uses the results to chose optimization passes and directly reduce code size~\cite{feedback}. \citet{serving} use LLMs to aid in searching the space of existing compiler optimizations in the GPU context. Our approach shares the integration of LLM-based reasoning with existing compiler optimizations, but differs in that it incorporates calls to compiler components with direct LLM code generation. We also allow the guiding agent to chose between the front-, middle-, and back-end of a compiler, rather than focusing on middle-end optimization passes.

\sumpara{Compiler Phase Ordering}
In exploring the space of possible rewrites and lowerings provided by existing LLVM and level-specific optimization agents, \tool's guiding agent confronts the well-known compiler phase ordering problem. Phase ordering followed soon after the first compiler with multiple phases~\cite{phase-micro, phase-global}, grappling with the issues of both synergies and anti-synergies between optimizations as does \tool. Subsequent approaches developed methods for conducting an exhaustive search of optimization orders~\cite{phase-exhaustive, phase-exhaustive-2} and for tailoring compiler phase choices to particular programs with iterative search~\cite{phase-graph, phase-iterative}. Phase ordering has also been tackled by pre-LLM machine learning approaches, including neural networks~\cite{micomp}, reinforcement learning~\cite{autophase, autophase2, phase-rl}, and others~\cite{recommender, boca}. We address phase ordering by empowering an LLM-based guiding agent to chose between existing compiler-provided rewrites and LLM-based level-specific optimization agents.

\section{Conclusion}
\label{sec:conclusion}
This work has presented LLM-compiler cooperation, a novel approach to LLM-based code optimization which integrates LLM-based code generation with existing compiler components at three different levels of abstraction: source code, intermediate representation, and low-level assembly. LLM-compiler cooperation harnesses the ``creativity'' of LLM-based optimization but tempers it with the guaranteed correctness of an existing compiler. We realize LLM-compiler cooperation as \tool, which integrates calls to LLM-based optimization agents with existing compiler components. Our approach combines LLM-based optimization at multiple levels of abstraction with a guiding agent that coordinates use of LLM-based and existing compiler-based optimization techniques. Our extensive evaluation shows that compiler-LLM cooperation outperforms both existing compiler optimizations and LLM-based optimization at any single level of abstraction, achieving an overall speedup of $1.25\times$ on average, with orders of magnitude improvement on individual input programs. 

%. We find that the guiding agent regularly makes use of all level-specific agents and existing compiler components, backtracking and providing feedback to improve performance on each input program, and 

\bibliographystyle{ACM-Reference-Format}
\bibliography{_references}

@article{compcert,
author = {Leroy, Xavier},
title = {Formal verification of a realistic compiler},
year = {2009},
issue_date = {July 2009},
publisher = {Association for Computing Machinery},
address = {New York, NY, USA},
volume = {52},
number = {7},
issn = {0001-0782},
url = {https://doi.org/10.1145/1538788.1538814},
doi = {10.1145/1538788.1538814},
journal = {Commun. ACM},
month = jul,
pages = {107–115},
numpages = {9}
}

@inproceedings{emi,
author = {Le, Vu and Afshari, Mehrdad and Su, Zhendong},
title = {Compiler validation via equivalence modulo inputs},
year = {2014},
isbn = {9781450327848},
publisher = {Association for Computing Machinery},
address = {New York, NY, USA},
url = {https://doi.org/10.1145/2594291.2594334},
doi = {10.1145/2594291.2594334},
booktitle = {Proceedings of the 35th ACM SIGPLAN Conference on Programming Language Design and Implementation},
pages = {216–226},
numpages = {11},
location = {Edinburgh, United Kingdom},
series = {PLDI '14}
}

@misc{scaling,
      title={Inference-Time Scaling for Complex Tasks: Where We Stand and What Lies Ahead}, 
      author={Vidhisha Balachandran and Jingya Chen and Lingjiao Chen and Shivam Garg and Neel Joshi and Yash Lara and John Langford and Besmira Nushi and Vibhav Vineet and Yue Wu and Safoora Yousefi},
      year={2025},
      eprint={2504.00294},
      archivePrefix={arXiv},
      primaryClass={cs.LG},
      url={https://arxiv.org/abs/2504.00294}, 
}

@inproceedings{
toolLLM,
title={Tool{LLM}: Facilitating Large Language Models to Master 16000+ Real-world {API}s},
author={Yujia Qin and Shihao Liang and Yining Ye and Kunlun Zhu and Lan Yan and Yaxi Lu and Yankai Lin and Xin Cong and Xiangru Tang and Bill Qian and Sihan Zhao and Lauren Hong and Runchu Tian and Ruobing Xie and Jie Zhou and Mark Gerstein and dahai li and Zhiyuan Liu and Maosong Sun},
booktitle={The Twelfth International Conference on Learning Representations},
year={2024},
url={https://openreview.net/forum?id=dHng2O0Jjr}
}

@misc{nl-tools,
      title={Natural Language Tools: A Natural Language Approach to Tool Calling In Large Language Agents}, 
      author={Reid T. Johnson and Michelle D. Pain and Jordan D. West},
      year={2025},
      eprint={2510.14453},
      archivePrefix={arXiv},
      primaryClass={cs.CL},
      url={https://arxiv.org/abs/2510.14453}, 
}

@inproceedings{llmDecomp,
   title={LLM4Decompile: Decompiling Binary Code with Large Language Models},
   url={http://dx.doi.org/10.18653/v1/2024.emnlp-main.203},
   DOI={10.18653/v1/2024.emnlp-main.203},
   booktitle={Proceedings of the 2024 Conference on Empirical Methods in Natural Language Processing},
   publisher={Association for Computational Linguistics},
   author={Tan, Hanzhuo and Luo, Qi and Li, Jing and Zhang, Yuqun},
   year={2024},
   pages={3473–3487} }

@inproceedings{tv-necula,
author = {Necula, George C.},
title = {Translation validation for an optimizing compiler},
year = {2000},
isbn = {1581131992},
publisher = {Association for Computing Machinery},
address = {New York, NY, USA},
url = {https://doi.org/10.1145/349299.349314},
doi = {10.1145/349299.349314},
booktitle = {Proceedings of the ACM SIGPLAN 2000 Conference on Programming Language Design and Implementation},
pages = {83–94},
numpages = {12},
location = {Vancouver, British Columbia, Canada},
series = {PLDI '00}
}

@inproceedings{reward-hacking,
author = {Skalse, Joar and Howe, Nikolaus H. R. and Krasheninnikov, Dmitrii and Krueger, David},
title = {Defining and characterizing reward hacking},
year = {2022},
isbn = {9781713871088},
publisher = {Curran Associates Inc.},
address = {Red Hook, NY, USA},
booktitle = {Proceedings of the 36th International Conference on Neural Information Processing Systems},
articleno = {687},
numpages = {12},
location = {New Orleans, LA, USA},
series = {NIPS '22}
}

@inproceedings{alive2,
author = {Lopes, Nuno P. and Lee, Juneyoung and Hur, Chung-Kil and Liu, Zhengyang and Regehr, John},
title = {Alive2: bounded translation validation for LLVM},
year = {2021},
isbn = {9781450383912},
publisher = {Association for Computing Machinery},
address = {New York, NY, USA},
url = {https://doi.org/10.1145/3453483.3454030},
doi = {10.1145/3453483.3454030},
booktitle = {Proceedings of the 42nd ACM SIGPLAN International Conference on Programming Language Design and Implementation},
pages = {65–79},
numpages = {15},
location = {Virtual, Canada},
series = {PLDI 2021}
}

@book{gcc,
author = {Stallman, Richard M. and GCC DeveloperCommunity},
title = {Using The Gnu Compiler Collection: A Gnu Manual For Gcc Version 4.3.3},
year = {2009},
isbn = {144141276X},
publisher = {CreateSpace},
address = {Scotts Valley, CA}
}

@inproceedings{llvm,
author = {Lattner, Chris and Adve, Vikram},
title = {LLVM: A Compilation Framework for Lifelong Program Analysis \& Transformation},
year = {2004},
isbn = {0769521029},
publisher = {IEEE Computer Society},
address = {USA},
booktitle = {Proceedings of the International Symposium on Code Generation and Optimization: Feedback-Directed and Runtime Optimization},
pages = {75},
location = {Palo Alto, California},
series = {CGO '04}
}

@inproceedings{
autophase2,
title={Autophase V2: Towards Function Level Phase Ordering Optimization},
author={Mohammed Almakki and Ayman Izzeldin and Qijing Huang and Ameer Haj Ali and Chris Cummins},
booktitle={Machine Learning for Computer Architecture and Systems 2022},
year={2022},
url={https://openreview.net/forum?id=HEWao_L2IP_}
}

@inproceedings{phase-rl,
author = {Li, Zujie and Qin, Huabiao and Xie, Yixiang},
title = {Applying Knowledge-Guided Deep Reinforcement Learning with Graph Neural Networks for Compiler Optimization},
year = {2025},
isbn = {9798400712647},
publisher = {Association for Computing Machinery},
address = {New York, NY, USA},
url = {https://doi.org/10.1145/3727648.3727682},
doi = {10.1145/3727648.3727682},
booktitle = {Proceedings of the 4th International Conference on Computer, Artificial Intelligence and Control Engineering},
pages = {193–198},
numpages = {6},
location = {
},
series = {CAICE '25}
}

@inproceedings{boca,
author = {Chen, Junjie and Xu, Ningxin and Chen, Peiqi and Zhang, Hongyu},
title = {Efficient Compiler Autotuning via Bayesian Optimization},
year = {2021},
isbn = {9781450390859},
publisher = {IEEE Press},
url = {https://doi.org/10.1109/ICSE43902.2021.00110},
doi = {10.1109/ICSE43902.2021.00110},
booktitle = {Proceedings of the 43rd International Conference on Software Engineering},
pages = {1198–1209},
numpages = {12},
location = {Madrid, Spain},
series = {ICSE '21}
}

@inproceedings{recommender,
author = {Cereda, Stefano and Palermo, Gianluca and Cremonesi, Paolo and Doni, Stefano},
title = {A Collaborative Filtering Approach for the Automatic Tuning of Compiler Optimisations},
year = {2020},
isbn = {9781450370943},
publisher = {Association for Computing Machinery},
address = {New York, NY, USA},
url = {https://doi.org/10.1145/3372799.3394361},
doi = {10.1145/3372799.3394361},
booktitle = {The 21st ACM SIGPLAN/SIGBED Conference on Languages, Compilers, and Tools for Embedded Systems},
pages = {15–25},
numpages = {11},
location = {London, United Kingdom},
series = {LCTES '20}
}

@article{micomp,
author = {Ashouri, Amir H. and Bignoli, Andrea and Palermo, Gianluca and Silvano, Cristina and Kulkarni, Sameer and Cavazos, John},
title = {MiCOMP: Mitigating the Compiler Phase-Ordering Problem Using Optimization Sub-Sequences and Machine Learning},
year = {2017},
issue_date = {September 2017},
publisher = {Association for Computing Machinery},
address = {New York, NY, USA},
volume = {14},
number = {3},
issn = {1544-3566},
url = {https://doi.org/10.1145/3124452},
doi = {10.1145/3124452},
journal = {ACM Trans. Archit. Code Optim.},
month = sep,
articleno = {29},
numpages = {28}
}

@inproceedings{autophase,
 author = {Haj-Ali, Ameer and Huang, Qijing (Jenny) and Xiang, John and Moses, William and Asanovic, Krste and Wawrzynek, John and Stoica, Ion},
 booktitle = {Proceedings of Machine Learning and Systems},
 editor = {I. Dhillon and D. Papailiopoulos and V. Sze},
 pages = {70--81},
 title = {AutoPhase: Juggling HLS Phase Orderings in Random Forests with Deep Reinforcement Learning},
 url = {https://proceedings.mlsys.org/paper_files/paper/2020/file/5b47430e24a5a1f9fe21f0e8eb814131-Paper.pdf},
 volume = {2},
 year = {2020}
}

@article{phase-iterative,
author = {Purini, Suresh and Jain, Lakshya},
title = {Finding good optimization sequences covering program space},
year = {2013},
issue_date = {January 2013},
publisher = {Association for Computing Machinery},
address = {New York, NY, USA},
volume = {9},
number = {4},
issn = {1544-3566},
url = {https://doi.org/10.1145/2400682.2400715},
doi = {10.1145/2400682.2400715},
journal = {ACM Trans. Archit. Code Optim.},
month = jan,
articleno = {56},
numpages = {23}
}

@inproceedings{phase-graph,
author = {Nobre, Ricardo and Martins, Luiz G. A. and Cardoso, Jo\~{a}o M. P.},
title = {A graph-based iterative compiler pass selection and phase ordering approach},
year = {2016},
isbn = {9781450343169},
publisher = {Association for Computing Machinery},
address = {New York, NY, USA},
url = {https://doi.org/10.1145/2907950.2907959},
doi = {10.1145/2907950.2907959},
booktitle = {Proceedings of the 17th ACM SIGPLAN/SIGBED Conference on Languages, Compilers, Tools, and Theory for Embedded Systems},
pages = {21–30},
numpages = {10},
location = {Santa Barbara, CA, USA},
series = {LCTES 2016}
}

@article{phase-exhaustive-2,
author = {Kulkarni, Prasad A. and Whalley, David B. and Tyson, Gary S. and Davidson, Jack W.},
title = {Practical exhaustive optimization phase order exploration and evaluation},
year = {2009},
issue_date = {March 2009},
publisher = {Association for Computing Machinery},
address = {New York, NY, USA},
volume = {6},
number = {1},
issn = {1544-3566},
url = {https://doi.org/10.1145/1509864.1509865},
doi = {10.1145/1509864.1509865},
journal = {ACM Trans. Archit. Code Optim.},
month = apr,
articleno = {1},
numpages = {36}
}

@inproceedings{phase-exhaustive,
author = {Kulkarni, Prasad A. and Whalley, David B. and Tyson, Gary S. and Davidson, Jack W.},
title = {Exhaustive Optimization Phase Order Space Exploration},
year = {2006},
isbn = {0769524990},
publisher = {IEEE Computer Society},
address = {USA},
url = {https://doi.org/10.1109/CGO.2006.15},
doi = {10.1109/CGO.2006.15},
booktitle = {Proceedings of the International Symposium on Code Generation and Optimization},
pages = {306–318},
numpages = {13},
series = {CGO '06}
}

@article{phase-global,
author = {Benitez, M. E. and Davidson, J. W.},
title = {A portable global optimizer and linker},
year = {1988},
issue_date = {July 1988},
publisher = {Association for Computing Machinery},
address = {New York, NY, USA},
volume = {23},
number = {7},
issn = {0362-1340},
url = {https://doi.org/10.1145/960116.54023},
doi = {10.1145/960116.54023},
journal = {SIGPLAN Not.},
month = jun,
pages = {329–338},
numpages = {10}
}

@inproceedings{phase-micro,
author = {Vegdahl, Steven R.},
title = {Phase coupling and constant generation in an optimizing microcode compiler},
year = {1982},
publisher = {IEEE Press},
booktitle = {Proceedings of the 15th Annual Workshop on Microprogramming},
pages = {125–133},
numpages = {9},
location = {Palo Alto, California, USA},
series = {MICRO 15}
}

@misc{serving,
      title={REASONING COMPILER: LLM-Guided Optimizations for Efficient Model Serving}, 
      author={Sujun Tang and Christopher Priebe and Rohan Mahapatra and Lianhui Qin and Hadi Esmaeilzadeh},
      year={2025},
      eprint={2506.01374},
      archivePrefix={arXiv},
      primaryClass={cs.LG},
      url={https://arxiv.org/abs/2506.01374}, 
}

@inproceedings{triton,
author = {Tillet, Philippe and Kung, H. T. and Cox, David},
title = {Triton: an intermediate language and compiler for tiled neural network computations},
year = {2019},
isbn = {9781450367196},
publisher = {Association for Computing Machinery},
address = {New York, NY, USA},
url = {https://doi.org/10.1145/3315508.3329973},
doi = {10.1145/3315508.3329973},
booktitle = {Proceedings of the 3rd ACM SIGPLAN International Workshop on Machine Learning and Programming Languages},
pages = {10–19},
numpages = {10},
location = {Phoenix, AZ, USA},
series = {MAPL 2019}
}

@article{kernelbench,
  author       = {Anne Ouyang and
                  Simon Guo and
                  Simran Arora and
                  Alex L. Zhang and
                  William Hu and
                  Christopher R{\'{e}} and
                  Azalia Mirhoseini},
  title        = {KernelBench: Can LLMs Write Efficient {GPU} Kernels?},
  journal      = {CoRR},
  volume       = {abs/2502.10517},
  year         = {2025},
  url          = {https://doi.org/10.48550/arXiv.2502.10517},
  doi          = {10.48550/ARXIV.2502.10517},
  eprinttype    = {arXiv},
  eprint       = {2502.10517},
  bibsource    = {dblp computer science bibliography, https://dblp.org}
}

@article{assembly,
  author       = {Anjiang Wei and
                  Tarun Suresh and
                  Huanmi Tan and
                  Yinglun Xu and
                  Gagandeep Singh and
                  Ke Wang and
                  Alex Aiken},
  title        = {Improving Assembly Code Performance with Large Language Models via
                  Reinforcement Learning},
  journal      = {CoRR},
  volume       = {abs/2505.11480},
  year         = {2025},
  url          = {https://doi.org/10.48550/arXiv.2505.11480},
  doi          = {10.48550/ARXIV.2505.11480},
  eprinttype    = {arXiv},
  eprint       = {2505.11480},
  bibsource    = {dblp computer science bibliography, https://dblp.org}
}

@article{compileAgent,
  author       = {Li Hu and
                  Guoqiang Chen and
                  Xiuwei Shang and
                  Shaoyin Cheng and
                  Benlong Wu and
                  Gangyang Li and
                  Xu Zhu and
                  Weiming Zhang and
                  Nenghai Yu},
  title        = {CompileAgent: Automated Real-World Repo-Level Compilation with Tool-Integrated
                  LLM-based Agent System},
  journal      = {CoRR},
  volume       = {abs/2505.04254},
  year         = {2025},
  url          = {https://doi.org/10.48550/arXiv.2505.04254},
  doi          = {10.48550/ARXIV.2505.04254},
  eprinttype    = {arXiv},
  eprint       = {2505.04254},
  bibsource    = {dblp computer science bibliography, https://dblp.org}
}

@article{feedback,
  author       = {Dejan Grubisic and
                  Chris Cummins and
                  Volker Seeker and
                  Hugh Leather},
  title        = {Compiler generated feedback for Large Language Models},
  journal      = {CoRR},
  volume       = {abs/2403.14714},
  year         = {2024},
  url          = {https://doi.org/10.48550/arXiv.2403.14714},
  doi          = {10.48550/ARXIV.2403.14714},
  eprinttype    = {arXiv},
  eprint       = {2403.14714},
  bibsource    = {dblp computer science bibliography, https://dblp.org}
}

@inproceedings{llmCompiler,
author = {Cummins, Chris and Seeker, Volker and Grubisic, Dejan and Roziere, Baptiste and Gehring, Jonas and Synnaeve, Gabriel and Leather, Hugh},
title = {LLM Compiler: Foundation Language Models for Compiler Optimization},
year = {2025},
isbn = {9798400714078},
publisher = {Association for Computing Machinery},
address = {New York, NY, USA},
url = {https://doi.org/10.1145/3708493.3712691},
doi = {10.1145/3708493.3712691},
booktitle = {Proceedings of the 34th ACM SIGPLAN International Conference on Compiler Construction},
pages = {141–153},
numpages = {13},
location = {Las Vegas, NV, USA},
series = {CC '25}
}

@misc{kevin,
  title={Multi-Turn RL Training for CUDA Kernel Generation},
  author={Carlo Baronio and Pietro Marsella and Ben Pan and Silas Alberti},
  howpublished={https://cognition.ai/blog/kevin-32b}}

@misc{kernelLLM,
    title={KernelLLM: Making Kernel Development More Accessible},
    author={Fisches, Zacharias V. and Paliskara, Sahan and Guo, Simon and Zhang, Alex and Spisak, Joe and Cummins, Chris and Leather, Hugh and Synnaeve, Gabriel and Isaacson, Joe and Markosyan, Aram and Saroufim, Mark},
    year={2025},
    howpublished={https://huggingface.co/facebook/KernelLLM}
}

@misc{metrKernel,
    title = {Measuring Automated Kernel Engineering},
    author = {METR},
    howpublished = {\url{https://metr.org/blog/2025-02-14-measuring-automated-kernel-engineering/}},
    year = {2025},
    month = {02}
}

@misc{nvidiaKernel,
    title = {Automating GPU Kernel Generation with DeepSeek-R1 and Inference Time Scaling},
    author = {Chen, Terry and Xu, Bing and Devleker, Kirthi},
    howpublished = {https://developer.nvidia.com/blog/automating-gpu-kernel-generation-with-deepseek-r1-and-inference-time-scaling/},
    year = {2025},
    month = {02}
}

@inproceedings{codenet,
  author       = {Ruchir Puri and
                  David S. Kung and
                  Geert Janssen and
                  Wei Zhang and
                  Giacomo Domeniconi and
                  Vladimir Zolotov and
                  Julian Dolby and
                  Jie Chen and
                  Mihir R. Choudhury and
                  Lindsey Decker and
                  Veronika Thost and
                  Luca Buratti and
                  Saurabh Pujar and
                  Shyam Ramji and
                  Ulrich Finkler and
                  Susan Malaika and
                  Frederick Reiss},
  editor       = {Joaquin Vanschoren and
                  Sai{-}Kit Yeung},
  title        = {CodeNet: {A} Large-Scale {AI} for Code Dataset for Learning a Diversity
                  of Coding Tasks},
  booktitle    = {Proceedings of the Neural Information Processing Systems Track on
                  Datasets and Benchmarks 1, NeurIPS Datasets and Benchmarks 2021, December
                  2021, virtual},
  year         = {2021},
  url          = {https://datasets-benchmarks-proceedings.neurips.cc/paper/2021/hash/a5bfc9e07964f8dddeb95fc584cd965d-Abstract-round2.html},
  bibsource    = {dblp computer science bibliography, https://dblp.org}
}

@article{planAct,
  author       = {Lutfi Eren Erdogan and
                  Nicholas Lee and
                  Sehoon Kim and
                  Suhong Moon and
                  Hiroki Furuta and
                  Gopala Anumanchipalli and
                  Kurt Keutzer and
                  Amir Gholami},
  title        = {Plan-and-Act: Improving Planning of Agents for Long-Horizon Tasks},
  journal      = {CoRR},
  volume       = {abs/2503.09572},
  year         = {2025},
  url          = {https://doi.org/10.48550/arXiv.2503.09572},
  doi          = {10.48550/ARXIV.2503.09572},
  eprinttype    = {arXiv},
  eprint       = {2503.09572},
  bibsource    = {dblp computer science bibliography, https://dblp.org}
}

@inproceedings{selfRefine,
  author       = {Aman Madaan and
                  Niket Tandon and
                  Prakhar Gupta and
                  Skyler Hallinan and
                  Luyu Gao and
                  Sarah Wiegreffe and
                  Uri Alon and
                  Nouha Dziri and
                  Shrimai Prabhumoye and
                  Yiming Yang and
                  Shashank Gupta and
                  Bodhisattwa Prasad Majumder and
                  Katherine Hermann and
                  Sean Welleck and
                  Amir Yazdanbakhsh and
                  Peter Clark},
  editor       = {Alice Oh and
                  Tristan Naumann and
                  Amir Globerson and
                  Kate Saenko and
                  Moritz Hardt and
                  Sergey Levine},
  title        = {Self-Refine: Iterative Refinement with Self-Feedback},
  booktitle    = {Advances in Neural Information Processing Systems 36: Annual Conference
                  on Neural Information Processing Systems 2023, NeurIPS 2023, New Orleans,
                  LA, USA, December 10 - 16, 2023},
  year         = {2023},
  url          = {http://papers.nips.cc/paper\_files/paper/2023/hash/91edff07232fb1b55a505a9e9f6c0ff3-Abstract-Conference.html},
  bibsource    = {dblp computer science bibliography, https://dblp.org}
}

@inproceedings{searchBased,
  author       = {Shuzheng Gao and
                  Cuiyun Gao and
                  Wenchao Gu and
                  Michael R. Lyu},
  title        = {Search-Based LLMs for Code Optimization},
  booktitle    = {47th {IEEE/ACM} International Conference on Software Engineering,
                  {ICSE} 2025, Ottawa, ON, Canada, April 26 - May 6, 2025},
  pages        = {578--590},
  publisher    = {{IEEE}},
  year         = {2025},
  url          = {https://doi.org/10.1109/ICSE55347.2025.00021},
  doi          = {10.1109/ICSE55347.2025.00021},
  bibsource    = {dblp computer science bibliography, https://dblp.org}
}

@inproceedings{learningEdits,
  author       = {Alexander Shypula and
                  Aman Madaan and
                  Yimeng Zeng and
                  Uri Alon and
                  Jacob R. Gardner and
                  Yiming Yang and
                  Milad Hashemi and
                  Graham Neubig and
                  Parthasarathy Ranganathan and
                  Osbert Bastani and
                  Amir Yazdanbakhsh},
  title        = {Learning Performance-Improving Code Edits},
  booktitle    = {The Twelfth International Conference on Learning Representations,
                  {ICLR} 2024, Vienna, Austria, May 7-11, 2024},
  publisher    = {OpenReview.net},
  year         = {2024},
  url          = {https://openreview.net/forum?id=ix7rLVHXyY},
  bibsource    = {dblp computer science bibliography, https://dblp.org}
}

@article{rlef,
  author       = {Jonas Gehring and
                  Kunhao Zheng and
                  Jade Copet and
                  Vegard Mella and
                  Taco Cohen and
                  Gabriel Synnaeve},
  title        = {{RLEF:} Grounding Code LLMs in Execution Feedback with Reinforcement
                  Learning},
  journal      = {CoRR},
  volume       = {abs/2410.02089},
  year         = {2024},
  url          = {https://doi.org/10.48550/arXiv.2410.02089},
  doi          = {10.48550/ARXIV.2410.02089},
  eprinttype    = {arXiv},
  eprint       = {2410.02089},
  bibsource    = {dblp computer science bibliography, https://dblp.org}
}

@inproceedings{ecco,
  author       = {Siddhant Waghjale and
                  Vishruth Veerendranath and
                  Zhiruo Wang and
                  Daniel Fried},
  editor       = {Yaser Al{-}Onaizan and
                  Mohit Bansal and
                  Yun{-}Nung Chen},
  title        = {{ECCO:} Can We Improve Model-Generated Code Efficiency Without Sacrificing
                  Functional Correctness?},
  booktitle    = {Proceedings of the 2024 Conference on Empirical Methods in Natural
                  Language Processing, {EMNLP} 2024, Miami, FL, USA, November 12-16,
                  2024},
  pages        = {15362--15376},
  publisher    = {Association for Computational Linguistics},
  year         = {2024},
  url          = {https://doi.org/10.18653/v1/2024.emnlp-main.859},
  doi          = {10.18653/V1/2024.EMNLP-MAIN.859},
  bibsource    = {dblp computer science bibliography, https://dblp.org}
}

@article{r1-interpreter,
  author       = {Yongchao Chen and
                  Yueying Liu and
                  Junwei Zhou and
                  Yilun Hao and
                  Jingquan Wang and
                  Yang Zhang and
                  Chuchu Fan},
  title        = {R1-Code-Interpreter: Training LLMs to Reason with Code via Supervised
                  and Reinforcement Learning},
  journal      = {CoRR},
  volume       = {abs/2505.21668},
  year         = {2025},
  url          = {https://doi.org/10.48550/arXiv.2505.21668},
  doi          = {10.48550/ARXIV.2505.21668},
  eprinttype    = {arXiv},
  eprint       = {2505.21668},
  bibsource    = {dblp computer science bibliography, https://dblp.org}
}

@article{the-stack,
  author       = {Denis Kocetkov and
                  Raymond Li and
                  Loubna Ben Allal and
                  Jia Li and
                  Chenghao Mou and
                  Yacine Jernite and
                  Margaret Mitchell and
                  Carlos Mu{\~{n}}oz Ferrandis and
                  Sean Hughes and
                  Thomas Wolf and
                  Dzmitry Bahdanau and
                  Leandro von Werra and
                  Harm de Vries},
  title        = {The Stack: 3 {TB} of permissively licensed source code},
  journal      = {Trans. Mach. Learn. Res.},
  volume       = {2023},
  year         = {2023},
  url          = {https://openreview.net/forum?id=pxpbTdUEpD},
  bibsource    = {dblp computer science bibliography, https://dblp.org}
}

@inproceedings{lifting,
author = {Verbeek, Freek and Bockenek, Joshua and Fu, Zhoulai and Ravindran, Binoy},
title = {Formally verified lifting of C-compiled x86-64 binaries},
year = {2022},
isbn = {9781450392655},
publisher = {Association for Computing Machinery},
address = {New York, NY, USA},
url = {https://doi.org/10.1145/3519939.3523702},
doi = {10.1145/3519939.3523702},
booktitle = {Proceedings of the 43rd ACM SIGPLAN International Conference on Programming Language Design and Implementation},
pages = {934–949},
numpages = {16},
location = {San Diego, CA, USA},
series = {PLDI 2022}
}

@misc{strands,
  title={Introducing Strands Agents, an Open Source AI Agents SDK},
  author={CClare Liguori },
  howpublished={https://aws.amazon.com/blogs/opensource/introducing-strands-agents-an-open-source-ai-agents-sdk/},
  year={2025}
}

@misc{sampling,
      title={Large Language Monkeys: Scaling Inference Compute with Repeated Sampling}, 
      author={Bradley Brown and Jordan Juravsky and Ryan Ehrlich and Ronald Clark and Quoc V. Le and Christopher Ré and Azalia Mirhoseini},
      year={2024},
      eprint={2407.21787},
      archivePrefix={arXiv},
      primaryClass={cs.LG},
      url={https://arxiv.org/abs/2407.21787}, 
}

@inproceedings{tool-use,
  author       = {Timo Schick and
                  Jane Dwivedi{-}Yu and
                  Roberto Dess{\`{\i}} and
                  Roberta Raileanu and
                  Maria Lomeli and
                  Eric Hambro and
                  Luke Zettlemoyer and
                  Nicola Cancedda and
                  Thomas Scialom},
  editor       = {Alice Oh and
                  Tristan Naumann and
                  Amir Globerson and
                  Kate Saenko and
                  Moritz Hardt and
                  Sergey Levine},
  title        = {Toolformer: Language Models Can Teach Themselves to Use Tools},
  booktitle    = {Advances in Neural Information Processing Systems 36: Annual Conference
                  on Neural Information Processing Systems 2023, NeurIPS 2023, New Orleans,
                  LA, USA, December 10 - 16, 2023},
  year         = {2023},
  url          = {http://papers.nips.cc/paper\_files/paper/2023/hash/d842425e4bf79ba039352da0f658a906-Abstract-Conference.html},
  bibsource    = {dblp computer science bibliography, https://dblp.org}
}

@article{grammar-decoding,
  author       = {Kanghee Park and
                  Jiayu Wang and
                  Taylor Berg{-}Kirkpatrick and
                  Nadia Polikarpova and
                  Loris D'Antoni},
  title        = {Grammar-Aligned Decoding},
  journal      = {CoRR},
  volume       = {abs/2405.21047},
  year         = {2024},
  url          = {https://doi.org/10.48550/arXiv.2405.21047},
  doi          = {10.48550/ARXIV.2405.21047},
  eprinttype    = {arXiv},
  eprint       = {2405.21047},
  bibsource    = {dblp computer science bibliography, https://dblp.org}
}

@misc{xgrammar,
      title={XGrammar: Flexible and Efficient Structured Generation Engine for Large Language Models}, 
      author={Yixin Dong and Charlie F. Ruan and Yaxing Cai and Ruihang Lai and Ziyi Xu and Yilong Zhao and Tianqi Chen},
      year={2025},
      eprint={2411.15100},
      archivePrefix={arXiv},
      primaryClass={cs.CL},
      url={https://arxiv.org/abs/2411.15100}, 
}


\end{document}